\documentclass[a4paper, 12pt]{article}
\usepackage{amsmath,amsthm,amssymb,mathrsfs,graphicx,url}
\usepackage{fullpage}
\usepackage{multirow}
\usepackage[usenames]{color}
\usepackage{ifthen}
\usepackage{setspace}
\usepackage{caption}
\usepackage{subfigure}
\usepackage[square,sort,comma,numbers]{natbib}
\usepackage{caption}
\usepackage{float}
\usepackage{graphicx}
\usepackage{algorithm}
\usepackage[noend]{algpseudocode}
\usepackage{makecell}
\usepackage{multirow}
\usepackage[margin=1.0in]{geometry}
\usepackage{fancyhdr}
\usepackage{lipsum}
\usepackage{titlesec}
\usepackage{bm}

\theoremstyle{plain}

\theoremstyle{remark}

\theoremstyle{definition}

\theoremstyle{remark}

\theoremstyle{definition}

\renewcommand{\phi}{\varphi}

\title{ 
A Study on the Association between Maternal Childhood Trauma Exposure and Placental-fetal Stress Physiology during Pregnancy}
\author{Eileen Zhang}
\date{%
Department of Statistics, University of California, Irvine\\%
congz10@uci.edu
}

\begin{document}
\maketitle
\begin{center}
\textbf{Abstract}
\end{center}
\textbf{Background} It has been found that the effect of childhood trauma (CT) exposure may pass on to the next generation. Scientists have hypothesized that the association between CT exposure and placental-fetal stress physiology is the mechanism. A study was conducted to examine the hypothesis. \\
\textbf{Method} To examine the association between CT exposure and placental corticotrophin-releasing hormone (pCRH), linear mixed effect model and hierarchical Bayesian linear model were constructed. In Bayesian inference, by providing conditionally conjugate priors, Gibbs sampler was used to draw MCMC samples. Piecewise linear mixed effect model was conducted in order to adjust to the dramatic change of pCRH at around week 20 into pregnancy. Pearson residual, QQ, ACF and trace plots were used to justify the model adequacy. Likelihood ratio test and DIC were utilized to model selection.\\
\textbf{Results} The association between CT exposure and pCRH during pregnancy is obvious. The effect of CT exposure on pCRH varies dramatically over gestational age. Women with one childhood trauma would experience 11.9\% higher in pCRH towards the end of pregnancy than those without childhood trauma. The increase rate of pCRH after week 20 is almost four-fold larger than that before week 20. Frequentist and Bayesian inference produce similar results.\\
\textbf{Conclusion} The findings support the hypothesis that the effect of CT exposure on pCRH over GA exists. The effect changes dramatically at around week 20 into pregnancy. 

\section{Introduction}
Childhood trauma (CT) is the traumatic experience that happens to children around age 0-6. It has been indicated that such adverse experience may have an effect on the developing brain \cite{ncb}, the development of depression and anxiety disorders \cite{hn}. Trauma survivors may become vulnerable during the period of time when there is other stress or changes in their lives\cite{skl}. Realizing the widespread consequences of CT, several studies have been conducted. Among them, an interesting finding is that CT may be transmitted between generations and its intergenerational impact will exist \cite{skl}. The mechanism behind it remains unsolved. Research results in this area proposed that the traumatized parents may be functional unavailable for their infant, which resulted in the enhanced symptomatology with their child \cite{wm}. Moreover, through parents' potential traumatizing behavior, child trauma may pass on to the next generation \cite{fi}. All the findings indicate the difficulty in searching for the mechanism of CT transmission. 

It has been studied that the placental corticotrophin-releasing hormone (pCRH) is the key to communicating between the mother and the unborn child\cite{sb}. The concentration of pCRH is highly related to the fetal and infant health development\cite{inft}. The pCRH system serves as sensor, transducer and effector of the fetal brain development and peripheral systems \cite{fp}. Motivated by these findings, a novel biological pathway has been proposed by Moog, N.K. et al. \cite{sb} to explain the mechanism of CT transmission over generations. In their outstanding work, they built the hypothesis that, through the effect of maternal CT exposure on placental-fetal stress physiology, especially pCRH, the intergenerational transmission may take place during gestation\cite{sb}. Specifically, their study was conducted in a cohort of 295 pregnant women along with their CT exposure measurement. Linear mixed effect model and Bayesian piecewise linear models were implemented to show the association between maternal CT exposure and placental-fetal stress physiology. 

In this study, the dataset consists of pCRH concentrations along with the CT exposure measurement and other information is obtained after a sociodemographically-diverse cohort of 88 pregnant women. The key scientific questions are to study the effect of CT exposure on pCRH over gestation (using both frequentist and Bayesian inference) and realize the unneglectable change in the rate of change in pCRH after gestational age goes beyond 20. Motivated by these, this report aims to provide a solution of those questions.  

\section{Data Description and Statistical Methods}
\subsection{Observed data}
The dataset in this study was collected during a sociodemographically-diverse cohort of 88 pregnant women. To measure the CT exposure, Childhood Trauma Questionnaire with 28 items have been assessed. It covered the following five dimensions of childhood maltreatments: emotional abuse (EA), physical abuse (PA), sexual abuse (SA), emotional neglect (EN) and physical neglect (PN). CT-Sum is the total number of those traumas that a mother had during her childhood. Placental CRH (pCRH) concentrations were also measured in maternal blood collected during the course of gestation. Besides those, the following information was also collected from each woman: gestation age in weeks (GA), depression score based on a questionnaire by the Center for Epidemiological Studies (DCES), indicator of obstetric risk conditions (OB-risk), pre-pregnancy Body Mass Index (BMI), childhood socioeconomic score using a 15-item measure that characterizes distinct aspects of economic status during childhood (CSES) and number of previous pregnancies
(Parity).
\subsection{Scientific Questions}
The following scientific questions are to be considered:\\
SQ1: Under the framework of frequentist inference, study the effect of CT-Sum on pCRH as gestation (GA) changes while considering all the potential confounding factors.\\
SQ2: Repeat the analysis of SQ1 under the framework of Bayesian inference. \\
SQ3: With the prior information that ``the rate of change of pCRH changes remarkably around 20 weeks into pregnancy'', modify the models. 
\subsection{Statistical Methods}
\label{sm1}
\textbf{As for the preliminary data exploration}  The main objective in this section is to study pregnant women characteristics during their gestation and how pCRH changes for different level of CT-Sum over GA. Quantitative covariates are presented in their mean, range and skewness along with graphing techniques such as spaghetti plot, regression line plot and scatterplot; categorical variables are presented in their percentage of the total population. 

\textbf{As for SQ1} The strategies were based on the following principles: \\
\textbf{Linear mixed effect model}  Since the current dataset contains multiple observations per subject and is unbalanced (each individual was measured at different gestation age), linear mixed effect models were employed to study the effect of CT-Sum on pCRH over GA. The assumption is that we assume subjects are independent with each other and linear relationship between pCRH (or transformed form of pCRH) and CT-Sum exists. \\
\textbf{Transform of covariates} It has been indicated that there exists an approximately exponential increase rate of pCRH over GA\cite{ska}. Following the same strategies of previous work\cite{sb,gao2019isbi,gao2020,gaoclassify,chengim}, log-transformation of pCRH has been used in fitting models.
The model follows that 
\begin{equation}
\log(pCRH_{i})=X_i\bm{\beta}+Z_ib_i+\epsilon_{i},
\label{model2}
\end{equation}
where $i=1,2,\cdots, 88.$ The subscript $i$ denotes subject id. $\bm{\beta}$ and $X_i$ are the coefficients and design matrix for the fixed effect. $Z_i$ and $b_i$ are the design matrix and random slope for the random effects. $\epsilon_{i}$ are error term within subjects.\\
 \textbf{Variable Selection} To study the effect of CT-Sum on pCRH over GA, covariates CT-Sum, GA and the interaction between them were included in the linear mixed effect model. It has been found that there is strong association between the socioeconomic background and childhood abuse, which results in the influence on pCRH \cite{lth,gaometron}. Thus, we have strong evidence that CSES should be in the fixed effects. Other covariates are either psychological or biophysical factors, which may be also associated with CT-Sum or pCRH. For instance, studies show that childhood sexual abuse have strong impact on depression during pregnancy \cite{hc,gaoner2019}. Childhood trauma survivors may deny their pregnancy or hide it from others \cite{wsg,gaoregular,wangner2019}, which may result in low number of previous pregnancies. Thus, in this study, we include all the factors in the fixed effects of the model. To take care of the variability between subjects in the effect of CT-Sum on pCRH, preliminary analysis suggests that the intercept varies across subjects. Moreover, it is also shown that the change rate of $\log pCRH$ over GA may also vary across subjects. Hence, we consider to include both of random intercept and slope of GA in the model. Likelihood ratio tests suggests that the random slope may not be necessarily included. Analysis in more details can be found in section \ref{sqq1}. 
 
\textbf{As for SQ2}  Bayesian inference was implemented to study the effect of CT-Sum on pCRH over GA. The strategies are in the following principles:\\
\textbf{Bayesian hierarchical model} With the same argument in SQ1, model (\ref{model2}) was also implemented in this section. To address the question in Bayesian inference, we have the following prior 
\begin{align*}
&\tau_1^2 \sim \text{Inv}-\chi^2(c,d),
\sigma_\epsilon^2 \sim \text{Inv}-\chi^2(a,b),
\beta_l \sim N(0, \sigma_l^2), l=0, 1, \cdots, k,\\
& b_{1i}|\tau_1^2 \overset{iid}{\sim}N(0,\tau_1^2),
\epsilon_{ij}|\sigma_\epsilon^2 \overset{iid}{\sim}N(0,\sigma_\epsilon^2), i=1,2,\cdots, 88, j=1,2,\cdots,n_i.
\end{align*}
As is stated in SQ1, we involve intercept in the random effects, which is denoted as $b_{1i}$. $\beta_l$ denotes the coefficient of the covariates in the fixed effect.\\
\textbf{MCMC estimation} Gibbs sampler was implemented in making inference of parameters of interest. The derivation of conditional distribution of parameters can be found in Appendix A. About 20\% of MCMC samplers were burned in to make inference of parameters of interest. Point estimate of parameters was calculated by the sample average. To get the 95\% credible interval, sample quantiles were used to approximate the lower and upper bounds. Besides that, MCMC convergence diagnosis was also conducted.\\
\textbf{Variable selection} From previous arguments, we have involved all the factors and the interaction CT-Sum*GA in the fixed effects and intercept in the random effects. In order to determine whether to include random slope of GA, DIC (Deviance Information Criterion) has been used to compare those two models.

\textbf{As for SQ3}  Piecewise linear mixed effect model was employed in studying the dramatic change of rate at around 20 weeks. A knot of 20 was made in covariate GA. Based on the model proposed in SQ1, we assume additional slope and intercept when GA is larger than 20. Likelihood ratio test was conducted to compare models with only additional slope and with both of additional slope and intercept. 

\subsection{Model Diagnosis}
To evaluate the model adequacy proposed in SQ1 and SQ3, Pearson residual plots have been employed to justify the mean model. QQ plots were also implemented to check the normality assumption. To justify the model proposed in SQ2, we employed statistics $T(y,\theta)=-2\sum_{i=1}^N\log(p(y_i|\theta))$. Based on the posterior samplers, data $y^{rep}$ is generated. We calculate the predictive p-value as $pB=Pr(T(y^{rep},\theta)>T(y,\theta)).$ If the value is extremely small, there may be some trouble in model adequacy.

\section{Result}
\subsection{Exploratory Data Analysis}
\label{explanatory}
Summary of descriptive statistics of categorical variables is shown in Table \ref{ds1} (Appendix A). It can be found that almost half of the subjects have no childhood trauma experience. The number of those who experience one to three childhood traumas are roughly the same (17.0\%, 14.8\% and 12.5\%). Only 4.5\% of the subjects have 4 childhood traumas. Most of the pregnant women (68.2\%) have low obstetric risk, which compares to 31.8\% as the high obstetric risk. 39.8\% of the women have no previous pregnancy and 38.6\% of them have one before. Only a small portion of subjects (12.5\%) have two previous pregnancies, along with 6.8\% have three and 2.3\% have four previous pregnancies. Table \ref{ds2} (Appendix A) presents characteristic of quantitative variables in this study. The sample average depression score is 0.65. The pre-pregnancy body mass index is 24.56 on average and 11.50 is the average childhood socioeconomic score among all the subjects. The gestational age is roughly centered around 26.73 weeks. The placental corticotrophin-releasing hormone has the average of 236.80. Three variables (DECS, CSES and GA) are not
highly skewed (skewness is 0.88, -0.77 and 0.01 respectively). BMI and the response variable pCRH are highly skewed. 

\begin{figure}[H] \centering
                 \begin{tabular}{cc}
    \includegraphics[width=.4\textwidth]{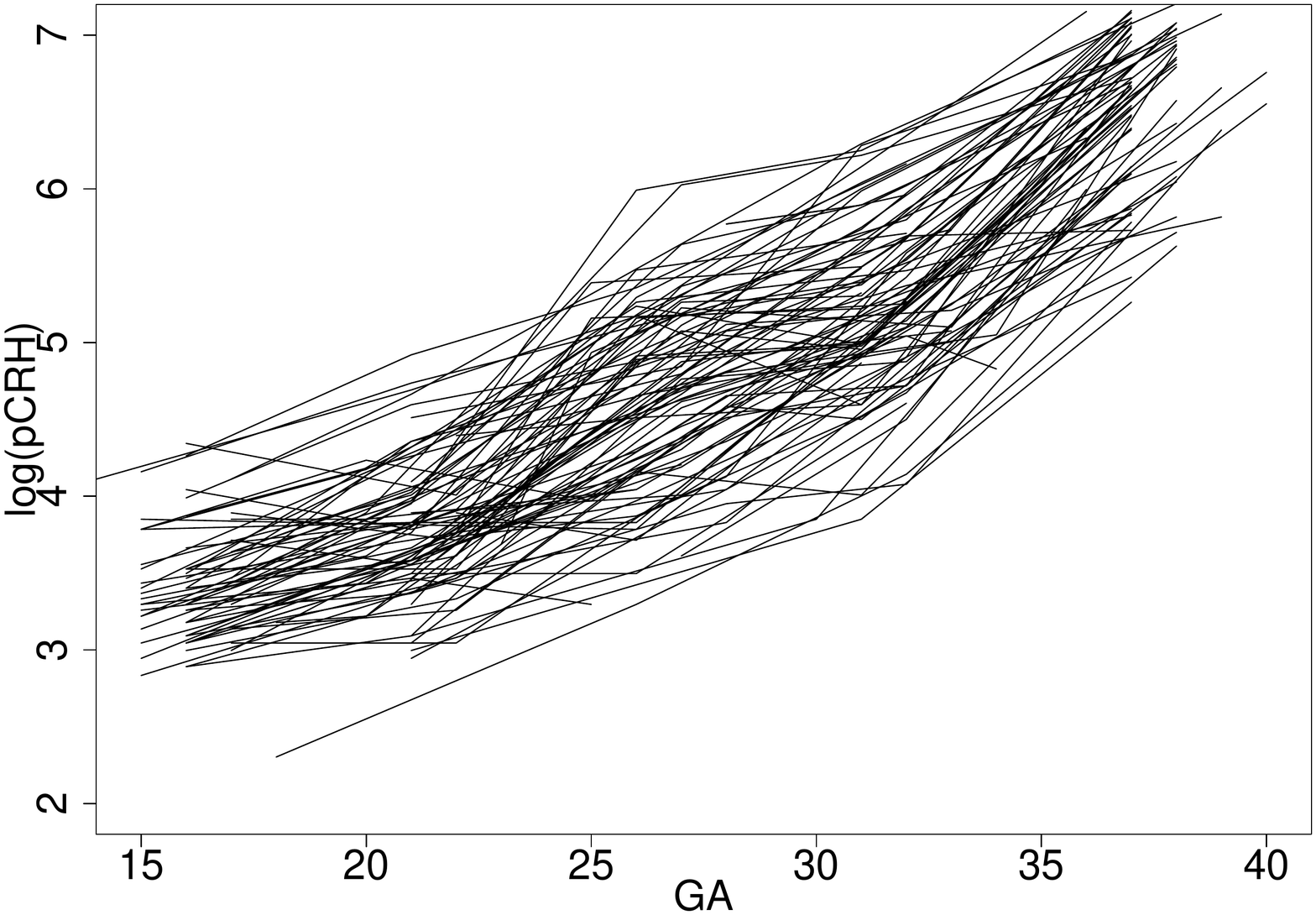} &  \includegraphics[width=.4\textwidth]{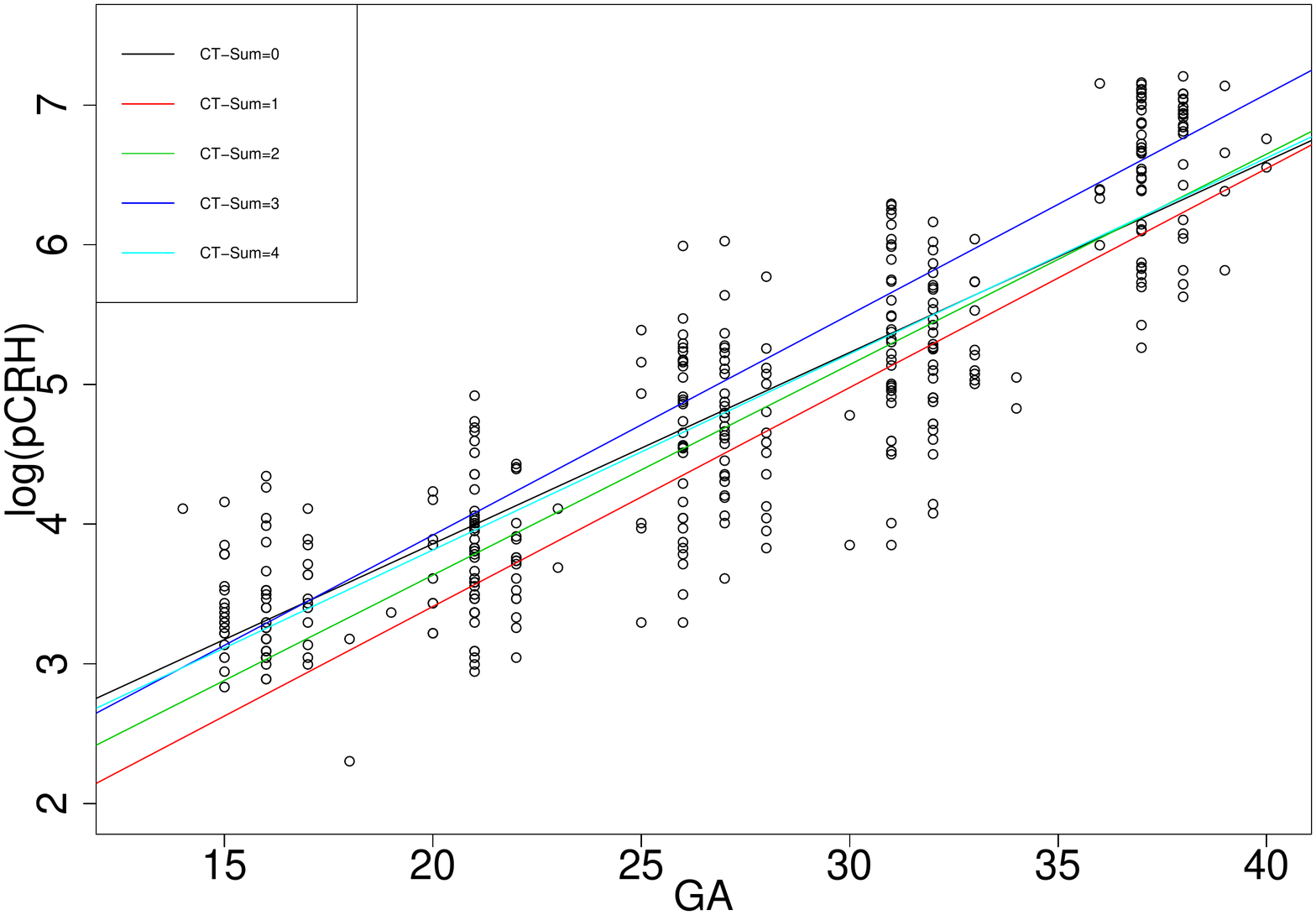}\\
    (A) Spaghetti plot of $\log pCRH$ & (B) Scatter plot overlaid with \\
    over GA for each individual  & regression lines for each CT-Sum\\
    \end{tabular}
 \caption{$\log pCRH$ over GA}
                               \label{tra}
\end{figure}

\begin{figure}[H] \centering
                 \begin{tabular}{cc}
    \includegraphics[width=.4\textwidth]{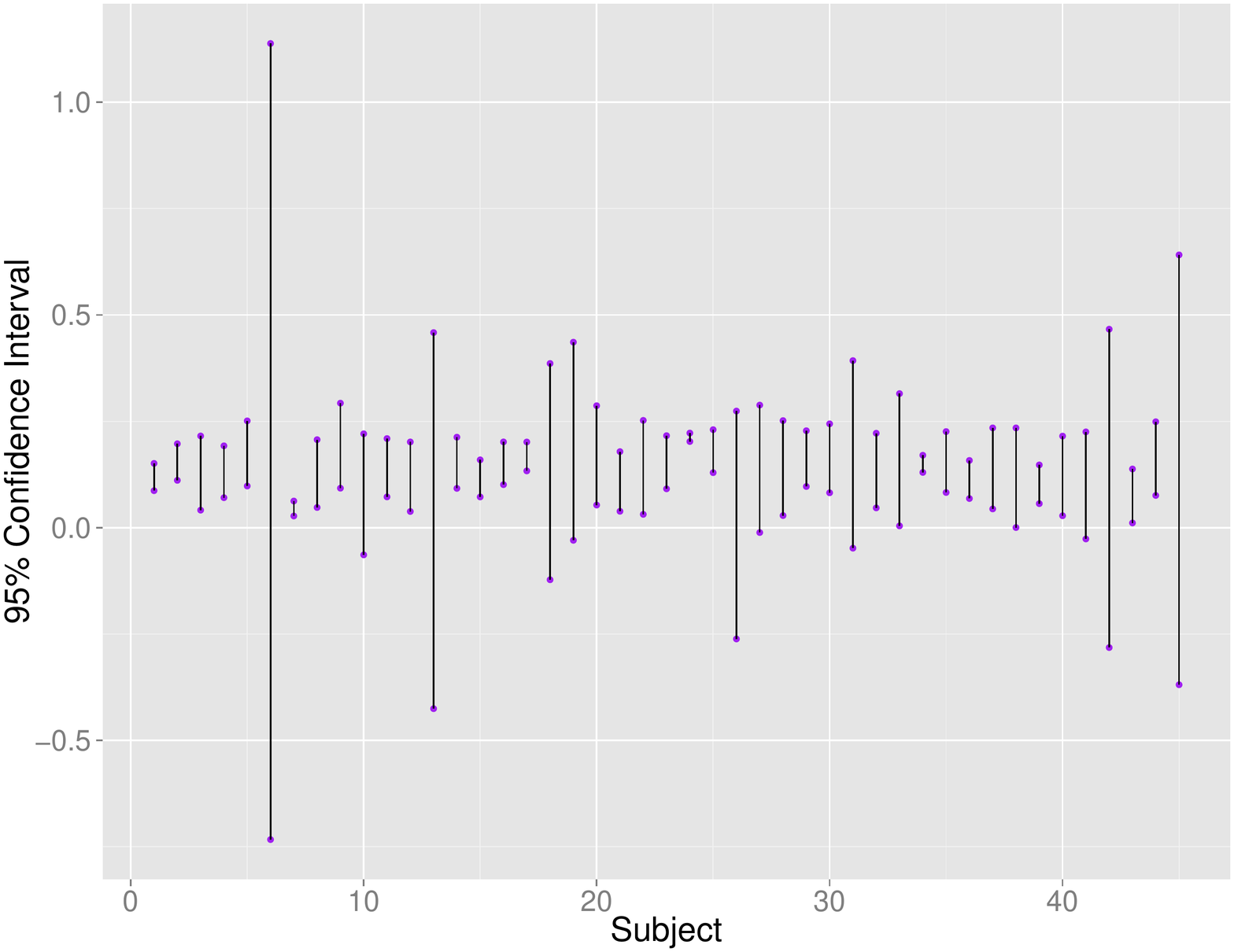} &  \includegraphics[width=.4\textwidth]{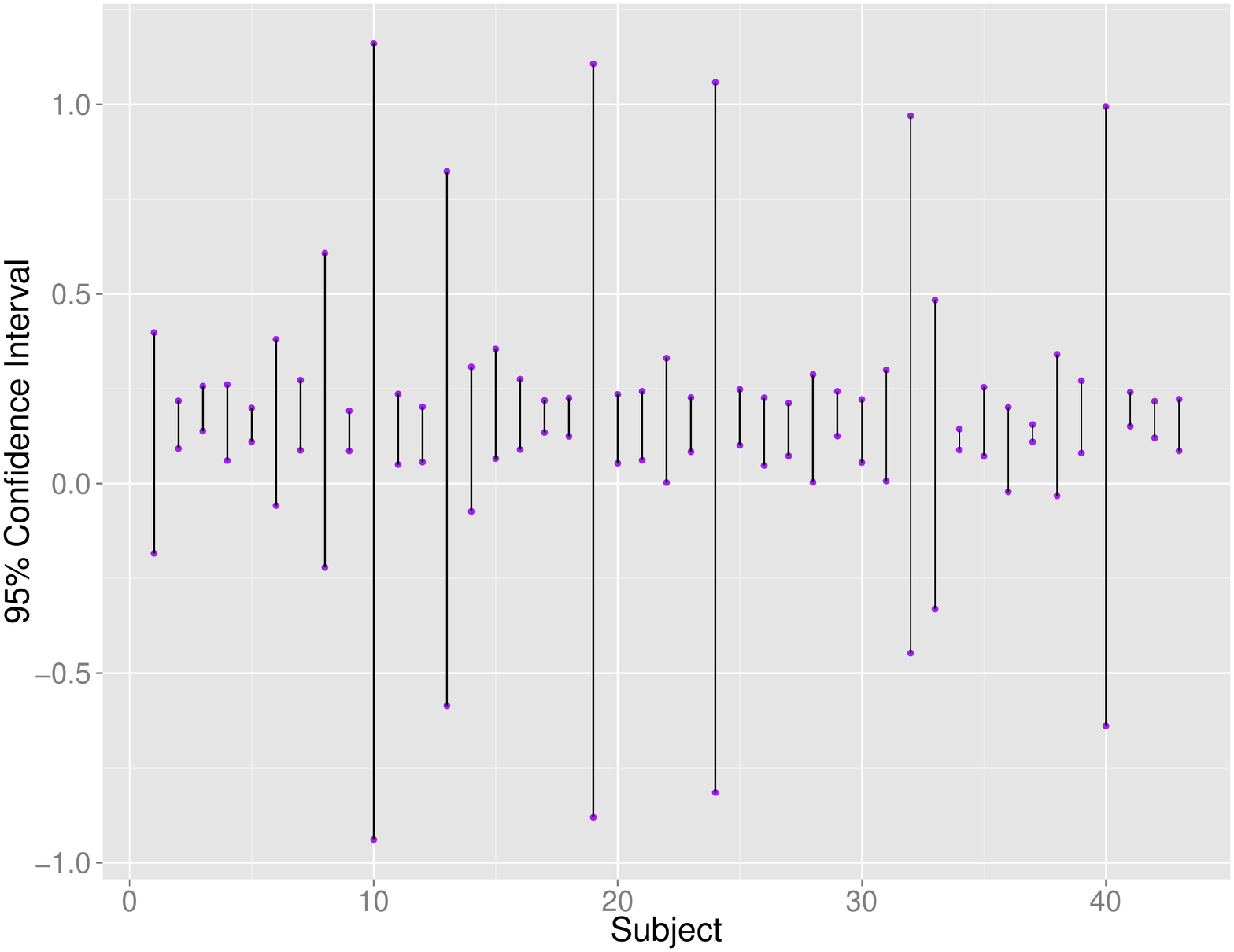}\\
    (A) $95\%$ confidence interval of slopes of GA & (B) $95\%$ confidence interval of slopes of GA \\
    for individuals without childhood trauma  &  for individuals with childhood trauma\\
    \end{tabular}
 \caption{ $95\%$ confidence interval of slopes of GA}
                               \label{tra1}
\end{figure}

\begin{figure}[H] \centering
                 \begin{tabular}{cc}
    \includegraphics[width=.4\textwidth]{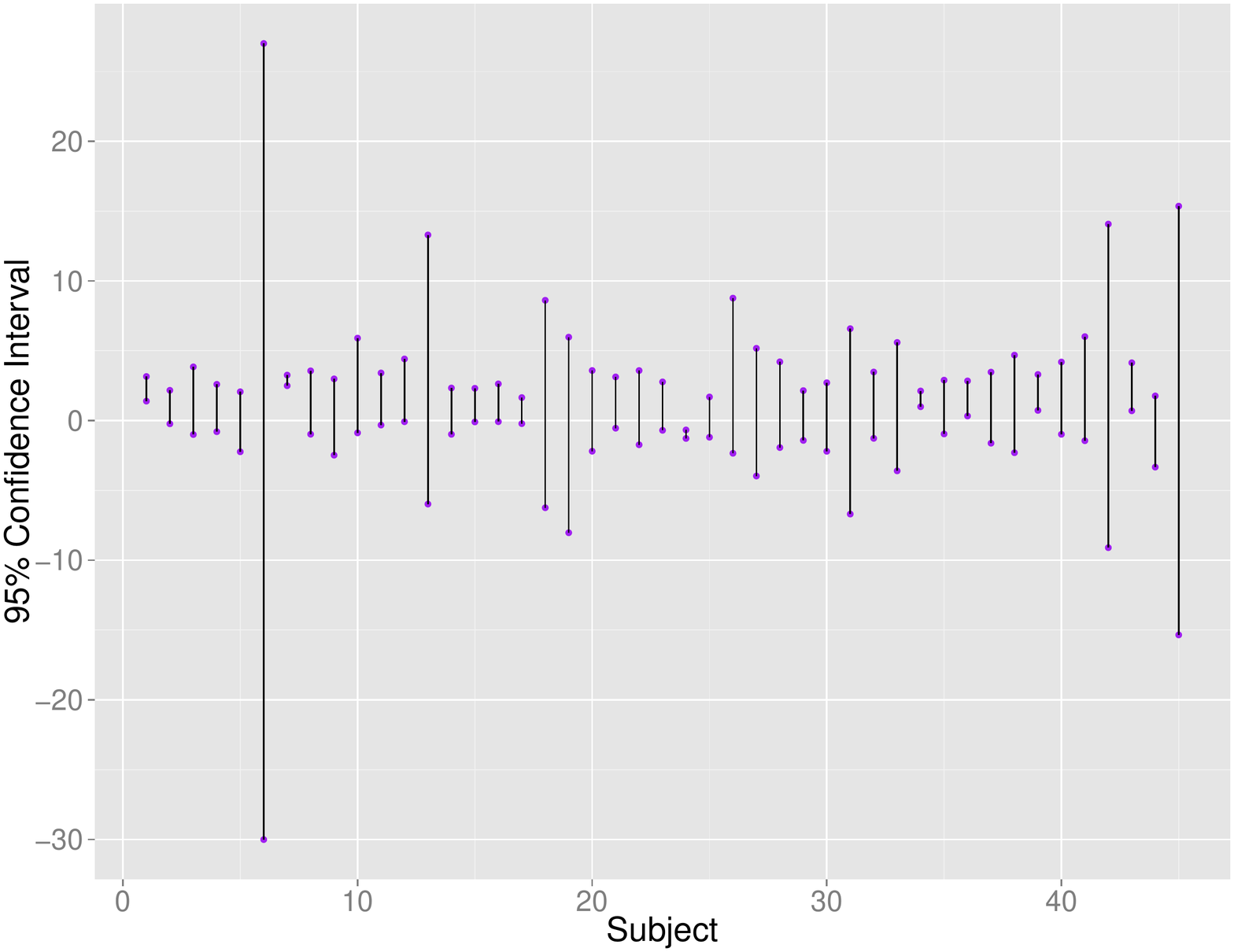} &  \includegraphics[width=.4\textwidth]{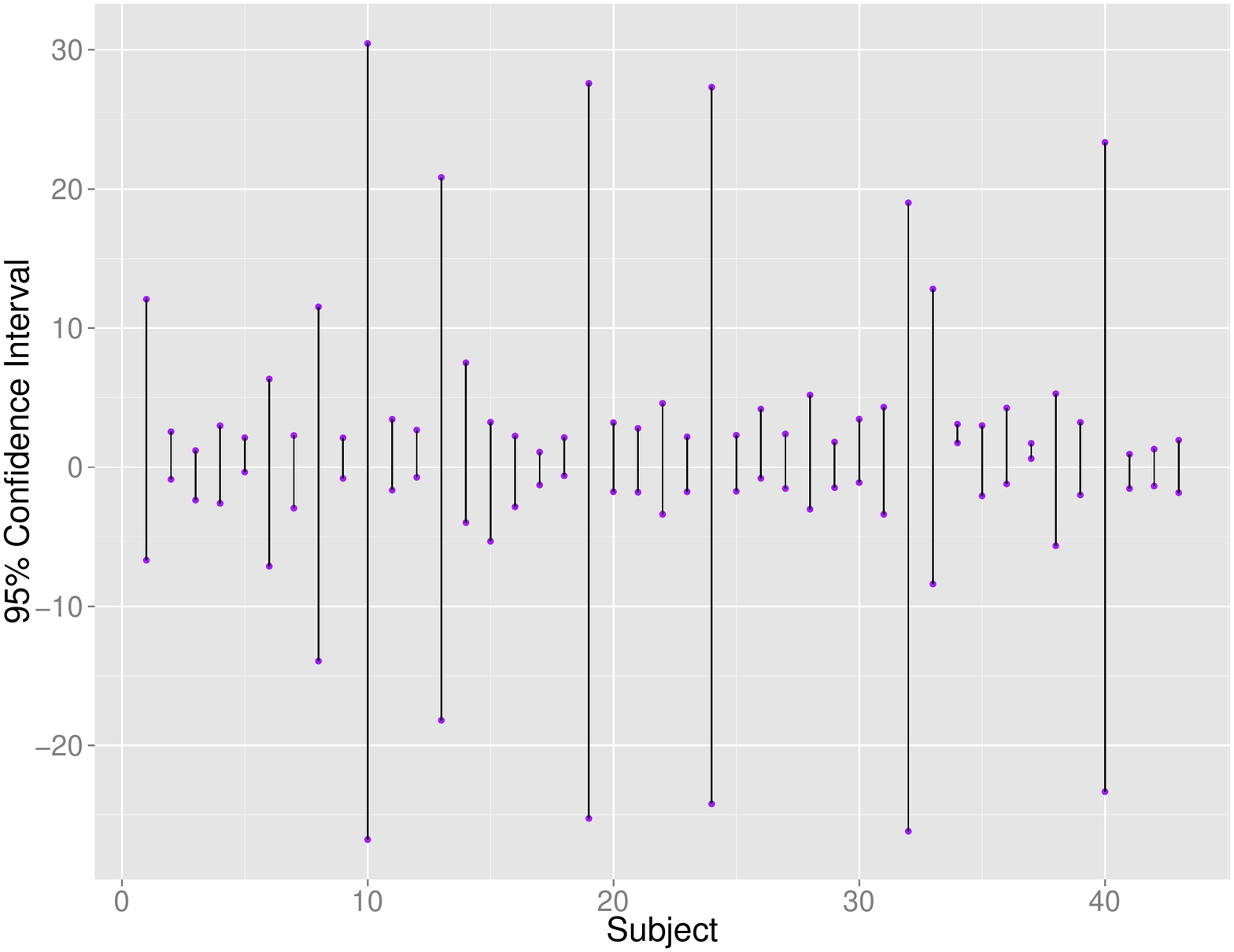}\\
    (A) $95\%$ confidence interval of intercepts & (B) $95\%$ confidence interval of intercepts \\
    for individuals without childhood trauma  &  for individuals with childhood trauma\\
    \end{tabular}
 \caption{ $95\%$ confidence interval of intercepts}
                               \label{tra2}
\end{figure}

To further study the effect of CT-Sum on $\log pCRH$ over GA, we conducted preliminary analysis. Figure \ref{tra}(A) shows the spaghetti plots of $\log pCRH$ over GA. It can be found that as GA develops, $\log pCRH$ increase linearly and the change of rates starts to grow larger when GA is around 25-30 week. After we control for the factor CT-Sum, the spaghetti plots are summarized in Figure \ref{trajc} (Appendix B). It shows that as GA closes to 15, the value of $\log pCRH$ for each individual is not consistent with each other after we control for the level of CT-Sum. The same pattern can also be found in Figure \ref{tra}(A). Figure \ref{tra2} shows the $95\%$ confidence intervals of the intercept of the fitted linear regression lines for each subject. It can be seen that after controlling for the factor of CT-Sum, the fitted intercept of each subjects differs from each other and the overlapping is relatively small. These findings motivate us to consider a random intercept into the linear mixed effects model. To further investigate the rate of change of $\log pCRH$ over GA, Figure \ref{tra}(B) shows the scatter plot overlaid with regression line after controlling for CT-Sum. It can be seen that for each value of CT-Sum, $\log pCRH$ increases over GA and the rate of change (slope) differs among different levels of CT-Sum, which indicates the effect of CT-Sum on $\log pCRH$ over GA. Figure \ref{tra1} presents the $95\%$ confidence intervals of the fitted slopes of GA for different subjects after controlling for CT-Sum. It shows that the fitted slope changes dramatically among different subjects especially for the group without childhood trauma. These findings also suggest us to introduce a random slope of GA in the linear mixed effect model. We will discuss on this in more details in Section \ref{sqq1}.

\subsection{Scientific Question 1}
\label{sqq1}
Following the analysis in Section \ref{explanatory}, a linear mixed effect model was fitted to the dataset. We include all the factors and the interaction CT-Sum*GA in the fixed effects. For the random effects, we considered two scenarios: random intercept only and both of random intercept and slope of GA. The two priori models are
\begin{align}
\label{sq2}
\begin{split}
\log(\text{pCRH}_{ij})=&\beta_0+\beta_1*\text{GA}_{ij}+\beta_{2}*\text{CT-Sum}_{i}
+\beta_{3}*\text{GA}_{ij}*\text{CT-Sum}_{i}\\
&+\beta_4*\text{BMI}_{i}+\beta_5*\text{CSES}_{i}+\beta_6*\text{DCES}_i+\beta_7*\text{OB-risk}_i+\beta_8*\text{Parity}_i\\
& +b_{1i}+b_{2i}*\text{GA}_{ij}+\epsilon_{ij},
\end{split}
\end{align}
and 
\begin{align}
\label{sq3}
\begin{split}
\log(\text{pCRH}_{ij})=&\beta_0+\beta_1*\text{GA}_{ij}+\beta_{2}*\text{CT-Sum}_{i}
+\beta_{3}*\text{GA}_{ij}*\text{CT-Sum}_{i}\\
&+\beta_4*\text{BMI}_{i}+\beta_5*\text{CSES}_{i}+\beta_6*\text{DCES}_i+\beta_7*\text{OB-risk}_i+\beta_8*\text{Parity}_i\\
& +b_{1i}+\epsilon_{ij},
\end{split}
\end{align}

To further compare model \ref{sq2} with model \ref{sq3}, a likelihood ratio test was conducted. The $\chi^2$ statistics roughly follows a mixture of $\chi^2$ distributions, $\frac{1}{2}\chi^2(1)+\frac{1}{2}\chi^2(2).$ The result (p value is 0.78) shows that there is no strong evidence to include random slope of GA in the model. 

Thus, the proposed model follows that 
\begin{align}
\label{sq1}
\begin{split}
\log(\text{pCRH}_{ij})=&\beta_0+\beta_1*\text{GA}_{ij}+\beta_{2}*\text{CT-Sum}_{i}
+\beta_{3}*\text{GA}_{ij}*\text{CT-Sum}_{i}\\
&+\beta_4*\text{BMI}_{i}+\beta_5*\text{CSES}_{i}+\beta_6*\text{DCES}_i+\beta_7*\text{OB-risk}_i+\beta_8*\text{Parity}_i\\
& +b_{1i}+\epsilon_{ij},
\end{split}
\end{align}
where fixed effects contain all the factors and the interaction GA*CT-Sum; the random effect involves intercept.

Model diagnosis was conducted in checking the constant variance assumption and normality assumption. Pearson residual plot in Figure \ref{md1} (Appendix C) indicates that the constant variance assumption holds since there is not obvious pattern from residuals. QQ plot in Figure \ref{md1} indicates normal assumption holds since the graph is closed to a line with slope 1. 

Table \ref{sqf1} shows the fixed effect estimates of linear mixed effect model (\ref{sq1}). It can be found that covariates of GA, BMI and CT-Sum*GA contribute significantly to the explanation of log(pCRH) change. Although CT-Sum does not preform statistical significant, there is no reason to argue that CT-Sum is not a persuasive predictor. Also, previous work \cite{sb} has shown the significance of CT-Sum. Moreover, it is shown from Table \ref{sqf1} that the association between CT-Sum and pCRH increases over gestational age (GA) since the estimate of CT-Sum*GA is positive. 

We will exponentiate the estimate to interpret the results. It can be concluded from the table that at gestational age 14 (the first time point collected in the dataset), give all the other factors the same, women with 1 childhood trauma tend to have 1.8\% lower in pCRH than those without childhood trauma. However, as gestational age goes up, the median pCRH is expected to goes up too. For example, if at gestational age 40 (the last time point in the dataset), the expected median pCRH value will increase by 11.9\% if the childhood trauma goes up by 1.

In Summary, to address SQ1, we propose model (\ref{sq1}) as the final fitted model. The effect of CT-Sum on pCRH over GA varies dramatically. At early gestational age, the effects of CT-Sum will decrease pCRH and as gestational age goes on, that effect becomes positive. If a pregnant woman experiences 1.2 childhood trauma (the average from the dataset), the expected median pCRH value will be 14.4\% higher towards the end of gestation compared to those without childhood trauma. On the other hand, at gestational age 26.7 (the average from the dataset), women with one childhood trauma has 4.7\% higher median pCRH value than those who do not have childhood trauma.


\begin{table}[h]
\hspace*{-2.0cm}
\caption{Fixed effect estimates of linear mixed effect model (\ref{sq1})}
\label{sqf1}
\begin{tabular}{llllll}
\hline
Covariates of fixed effect  & Estimate & Standard error & Degree of freedom & T value & P value* \\ \hline
   Intercept    &  1.750   &    0.270               &  113.200 & 6.502 &  $<0.001$ \\
    GA & 0.142  &  0.004              & 301.900  & 37.254 & $<0.001$  \\
      CT-Sum  &  -0.088  &  0.071               &   375.900 &  -1.242 & 0.215  \\ 
      CT-Sum*GA      &  0.005 & 0.002             &   299.400   & 1.971 & 0.050\\ 
  BMI & -0.021 & 0.007 & 86.900 & -3.056 & 0.003 \\
  CSES & -0.018 & 0.015 & 84.700 & -1.233 & 0.221\\
  DCES & -0.093 & 0.102 & 87.100 & -0.914 & 0.363\\
  OB-risk & 0.035 & 0.089 & 86.900 & 0.391 & 0.697\\
  Parity & -0.076 & 0.041 & 90.400 & -1.830 & 0.071\\
 \hline
 \multicolumn{5}{l}{*The p value is based on Satterthwaite approximation.}
\end{tabular}
\end{table}

\subsection{Scientific Question 2}
Following the argument in section \ref{sm1}, Bayesian inference has been made to study the association between CT-Sum and pCRH over GA. By similar arguments in section \ref{sqq1}, we have involved all the factors and the interaction GA*CT-Sum in the fixed effects. DIC was compared to determine whether to include random slope of GA in the model. Results (random intercept and slope: 684.12; random intercept only: 577.75) suggests only random intercept included in the model. Thus, we propose model (\ref{sq1}) as the final hierarchical Bayesian linear model as well.

Model diagnosis was conducted in checking the convergence of MCMC and also the adequacy of the model. Trace and ACF plots of parameters can be found in Figures \ref{mdb1} and \ref{mdb2} (Appendix D). They reveal that all the MCMC samplers are in good mixing and independent. The Gibbs samplers provide good estimate of the posterior distribution for each parameter. Figure \ref{md4} (Appendix D) compares to observed average test statistics defined in section \ref{sm1} with the values obtained from the replicated samples. The estimated $pB$ is 0.58, which implies the model fits the data well.  

The fixed effect estimates of the proposed model can be found in Table \ref{sq}. Similar to the frequentist inference, it shows that intercept, GA and BMI are significant since the credible interval does not contain 0. Although the interaction GA*CT-Sum covers 0 in their credible interval, we conclude it is still significant since the lower bound is close to 0. CT-Sum does not preform statistically significant but we can not neglect it by the same argument in section \ref{sqq1}. To interpret the results, we will still exponentiate the estimates. The positive value of 0.005 suggests positive association between CT-Sum and pCRH as gestational age increases. To address SQ2, we conclude that the increment of CT-Sum results in decrease in pCRH at early gestational age. However, as gestational age becomes large, increasing the CT-Sum will lead to the increment of pCRH. The arguments are similar to the frequentist inference. If a pregnant woman has 1.2 childhood trauma (the average of the dataset), the expected median pCRH value will be 14.4\% higher at the end of the gestation compared to those without childhood trauma. Moreover, at the average gestational age 26.7, women with one childhood trauma will have pCRH value 4.7\% higher than those without childhood trauma.  
\begin{table}[h]
\centering
\caption{Fixed effect estimates of hierarchical Bayesian model (\ref{sq1})}
\label{sq}
\begin{tabular}{lll}
\hline
Covariates of fixed effect  & Posterior mean & 95\% Probability Interval  \\ \hline
   Intercept    &  1.751 &    (1.152, 2.328)          \\
    GA & 0.141  &  (0.132, 0.151)        \\
      CT-Sum  &  -0.088  & (-0.265, 0.088)               \\ 
      CT-Sum*GA      &  0.005 & (-0.001, 0.011)         \\ 
  BMI & -0.020 & (-0.035, -0.006)\\
  CSES & -0.018 & (-0.049, 0.013)\\
  DCES & -0.098 & (-0.313, 0.118)\\
  OB-risk & 0.037 &(-0.149, 0.225)\\
  Parity & -0.077 & (-0.164, 0.011) \\
 \hline
\end{tabular}
\end{table}


\subsection{Scientific Question 3}
To modify model (\ref{sq1}) with the additional information, we made a knot of 20 in covariate GA. The question is whether additional slope, or both of additional slope and intercept should be involved in the fit model. To address this question, models with only additional slope and both of additional slope and intercept have been fitted. Likelihood ratio test (pvalue is 0.43) suggests that there is no further information indicating the necessity of both of additional intercept and slope. Hence, we will propose the model with only additional slope that follows
\begin{align}
\label{sq33}
\begin{split}
\log(\text{pCRH}_{ij})&=\beta_0+\beta_1*\text{GA}_{ij}+\beta_{2}*\text{CT-Sum}_{i}
+\beta_{3}*\text{GA}_{ij}*\text{CT-Sum}_{i}+\beta_4*\text{BMI}_{i}\\&
+\beta_5*(\text{GA}_{ij}-20)_++\beta_6*\text{CSES}_{i}+\beta_7*\text{Parity}_{i}\\
&+\beta_8*\text{DCES}_{i}+\beta_9*\text{OB-risk}_{i}+b_{1i}+\epsilon_{ij},
\end{split}
\end{align}
where $(\text{GA}_{ij}-20)_+=max\{\text{GA}_{ij}-20, 0\}.$

Model diagnosis was conducted to justify the adequacy, assumption of constant variance and normality. Residual plot can be found in Figure \ref{md3} (Appendix D). There is no obvious pattern and the residuals scatters around 0. Also, the QQ plot indicates the validity of normal assumption.

Summary of fixed effect estimates of model (\ref{sq33}) is shown in Table \ref{sqff2}. Similar to the results in SQ1 and SQ2, CT-Sum are not statistically significant. Since the objective is to study the change of effect of CT-Sum on pCRH over GA, there is no reason to remove CT-Sum from the model. 

Following similar strategies in SQ1 and SQ2, we will exponentiate the estimate of coefficients to interpret the results. From Table \ref{sqff2}, it is revealed that around 20 gestational age, there is a dramatic increase of the change rate of pCRH over gestation. In particular, among women whose gestational age is less than 20 and experience 1.2 childhood trauma (the average from the dataset), the expected median pCRH will increase by 3.8\% per week. But after 20 weeks, those women will experience a 17.9\% increase of the expected median pCRH per week towards the end of gestation. On the other hand, among those women with 1 childhood trauma (the mode of the dataset), the expected median pCRH increases by 3.9\% per week before week 20, 17.8\% per week after week 20. But among those without childhood trauma, the expected median pCRH goes up by 3.3\% (before week 20) and 17.2\% (after week 20). It shows that although at around week 20, the increment of pCRH becomes dramatic, women without childhood trauma still remain lower pCRH increase rate compared to those with childhood trauma.  
\begin{table}[h]
\caption{Fixed effect estimates of linear mixed effect model (\ref{sq33})}
\label{sqff2}
\begin{tabular}{llllll}
\hline
Covariates of fixed effect  & Estimate & Standard error & Degree of freedom & T value & P value* \\ \hline
   Intercept    &  3.729   &    0.374               &  300.500 & 9.961 &  $<0.001$ \\
    GA & 0.032  &  0.015             & 298.600  & 2.126 & 0.034  \\
     $(\text{GA}-20)_+$ & 0.127  &  0.017             & 297.400  & 7.428 &  $<0.001$ \\
      CT-Sum  &  -0.107 &  0.067               &   371.700 &  -1.617 & 0.107  \\ 
      CT-Sum*GA      &  0.005 & 0.002             &   298.400   &  2.409 & 0.016\\ 
  BMI & -0.020 & 0.007 & 87.200 & -3.020 & 0.003 \\
  CSES & -0.020 & 0.015 & 85.200 &  -1.349 & 0.181 \\
  Parity & -0.078 & 0.041 & 90.200 & -1.916 & 0.059 \\
  DCES & -0.084 & 0.100 & 87.400 & -0.838 & 0.404\\
  OB-risk & 0.039 & 0.088 & 87.200 & 0.437 & 0.663\\
 \hline
 \multicolumn{5}{l}{*The p value is based on Satterthwaite approximation.}
\end{tabular}
\end{table}

\section{Discussion}
\subsection{Conclusion}
The objectives of this study were to examine the effect of maternal CT exposure on pCRH and modify the model to realize the difference before and after gestational age 20 (in week). 

In regarding to the first scientific question, linear mixed effect models have been implemented. Covariates of all the factors and the interaction GA*CT-Sum were chosen as fixed effects. Intercept was in the random effects. Results indicated that the association between CT exposure and pCRH varied over gestational age. During the first couple of weeks into pregnancy, women with childhood trauma were likely to have lower pCRH than those without childhood trauma. However, as gestational age moved on, those with childhood trauma experienced much higher increase rate of pCRH. At the end of pregnancy (GA=40), women with 1 childhood trauma have almost 14.4\% higher value in the expected median pCRH than those without childhood trauma. 

In regarding to the second scientific question, hierarchical Bayesian linear mixed effect model was implemented. By choosing conditionally conjugate priors, Gibbs sampler was employed to obtain samplers from the posterior distribution of parameters. The same model in SQ1 was proposed after comparing the DIC values. Results are similar to the frequentist inference. At early gestational age, women with more childhood trauma would experience lower pCRH value. As the pregnancy moves on, more exposure to childhood trauma lead to much higher increase rate of pCRH. At the average gestational age (GA=26.7), women with one childhood trauma have 4.7\% higher pCRH value than those without childhood trauma experience.   

In regarding to the last scientific question, piecewise linear model with knot at 20 week was conducted. The results indicated that after week 20 into pregnancy, the increase of pCRH over GA became more and more dramatic. The increase rate per week (after week 20) is almost four-fold larger than that before week 20. Women without childhood trauma still remain lower increase rate than those with one childhood trauma before and after week 20. 
\subsection{Limitations}
The main drawbacks of this study lie in the following two aspects. On one hand, the response (pCRH) on different subjects were not measured at the same time point (gestational age). It has been pointed out that the unbalanced structure may result in misspecification of the within-subject association over continuous time \cite{hw}. If the response are not missing completely at random, such misspecification may lead to biased estimates of the mean response\cite{ru}. On the other hand, there may be more potential confounding factors. For instance, characteristics such as race, ethnicity, drug use, alcohol in pregnancy and age are not considered in this study, which may result in  unexpected models. 

\newpage
\appendix
\section{Appendix A}
The derivation of conditional distributions in implementing Gibbs sampler.
\begin{align*}
\label{gibbs}
& b_{1i}|\bm{\beta}, \tau_1^2,\sigma_\epsilon^2 \sim N(\frac{\sum_{j=1}^{n_i}(\log(pCRH_{ij})-X_{ij}\bm{\beta})/ \sigma_\epsilon^2}{n_i/\sigma_\epsilon^2 +1/ \tau_1^2}, (n_i/\sigma_\epsilon^2 +1/ \tau_1^2)^{-1}),\\
&\bm{\beta}|\tau_1^2, b_{1i}, \sigma_\epsilon^2 \sim N((1/\sigma_\epsilon^2)((1/\sigma_\epsilon^2)*X^TX+\Lambda_0)^{-1}X^T\tilde{Y}, ((1/\sigma_\epsilon^2)*X^TX+\Lambda_0)^{-1}), \\
&\text{where} X= \begin{bmatrix}
X_{11}\\
X_{12} \\
\cdots \\
X_{88,n_{88}}\end{bmatrix},   \Lambda_0=\begin{bmatrix}
1/ \sigma_0^2 & 0 & 0 \\
0 & \cdots & 0 \\
0& 0& 1/ \sigma_k^2 \end{bmatrix} ,\\
&\tilde{Y}=(\log(pCRH_{11})-b_{11}, \cdots, \log(pCRH_{88,n_{88}})-b_{1,88})^T.\\
&\tau_1^2|\bm{\beta}, b_{1i},\sigma_\epsilon^2 \sim \text{Inv}-\chi^2(88+c,\frac{\sum_{i=1}^{88}b_{1i}^2+c*d}{88+c})\\
&\sigma_\epsilon^2|\bm{\beta}, \tau_1^2, b_{1i}  \\
&\sim \text{Inv}-\chi^2(\sum_{i=1}^{88}n_i+a,\frac{\sum_{i=1}^{88}\sum_{j=1}^{n_i}(\log(pCRH_{ij})-X_{ij}\bm{\beta}-b_{1i})^2+a*b}{\sum_{i=1}^{88}n_i+a}).\\
\end{align*}
\begin{table}[h]
\centering
\caption{Descriptive statistics of categorical variables}
\label{ds1}
\begin{tabular}{lll}
\hline
Variables & Number of observations & Percentage    \\ \hline
        No childhood trauma (CT-Sum=0)   &   45   &        51.1\%                     \\
       One childhood trauma (CT-Sum=1)   &    15  &           17.0\%                  \\
     Two childhood trauma (CT-Sum=2)       &   13   &            14.8\%                 \\ 
       Three childhood trauma (CT-Sum=3)       &   11   &            12.5\%                 \\ 
     Four childhood trauma (CT-Sum=4)      &   4   &                  4.5\%           \\ 
    High obstetric risk (OB-risk=1)      &   28  &                  31.8\%           \\ 
   Low obstetric risk (OB-risk=0)      &   60  &                  68.2\%           \\ 
        No previous pregnancy (Parity=0)    &   35  &           39.8\%                  \\ 
           One previous pregnancy (Parity=1)    &   34   &          38.6\%                   \\ 
        Two previous pregnancies (Parity=2)     &   11   &             12.5\%                \\ 
         Three previous pregnancies (Parity=3)       &   6   &          6.8\%                   \\ 
          Four previous pregnancies (Parity=4)     &   2   & 2.3\%                            \\ 
 \hline
\end{tabular}
\end{table}
\begin{table}[h]
\caption{Descriptive statistics of quantitative variables}
\label{ds2}
\centering
\begin{tabular}{llll}
\hline
Variables & Mean & Range & Skewness \\ \hline
      Depression score (DCES)   &  0.65                 &   1.72   &  0.88\\
       Pre-pregnancy body mass index (BMI)   &  24.56              &  30.00   & 1.15 \\
      Childhood socioeconomic score (CSES)    &   11.50             &   11.00    & -0.77  \\ 
      Gestational age (in weeks) (GA)    &   26.73           &   26.00    & 0.01 \\ 
       Placental corticotrophin-releasing hormone (pCRH) & 236.80 & 1337.00 & 1.94\\
 \hline
\end{tabular}
\end{table}

\newpage
\section{Appendix B}
\begin{figure}[h] \centering
                              \begin{tabular}{cc}
                               \includegraphics[width=.4\textwidth]{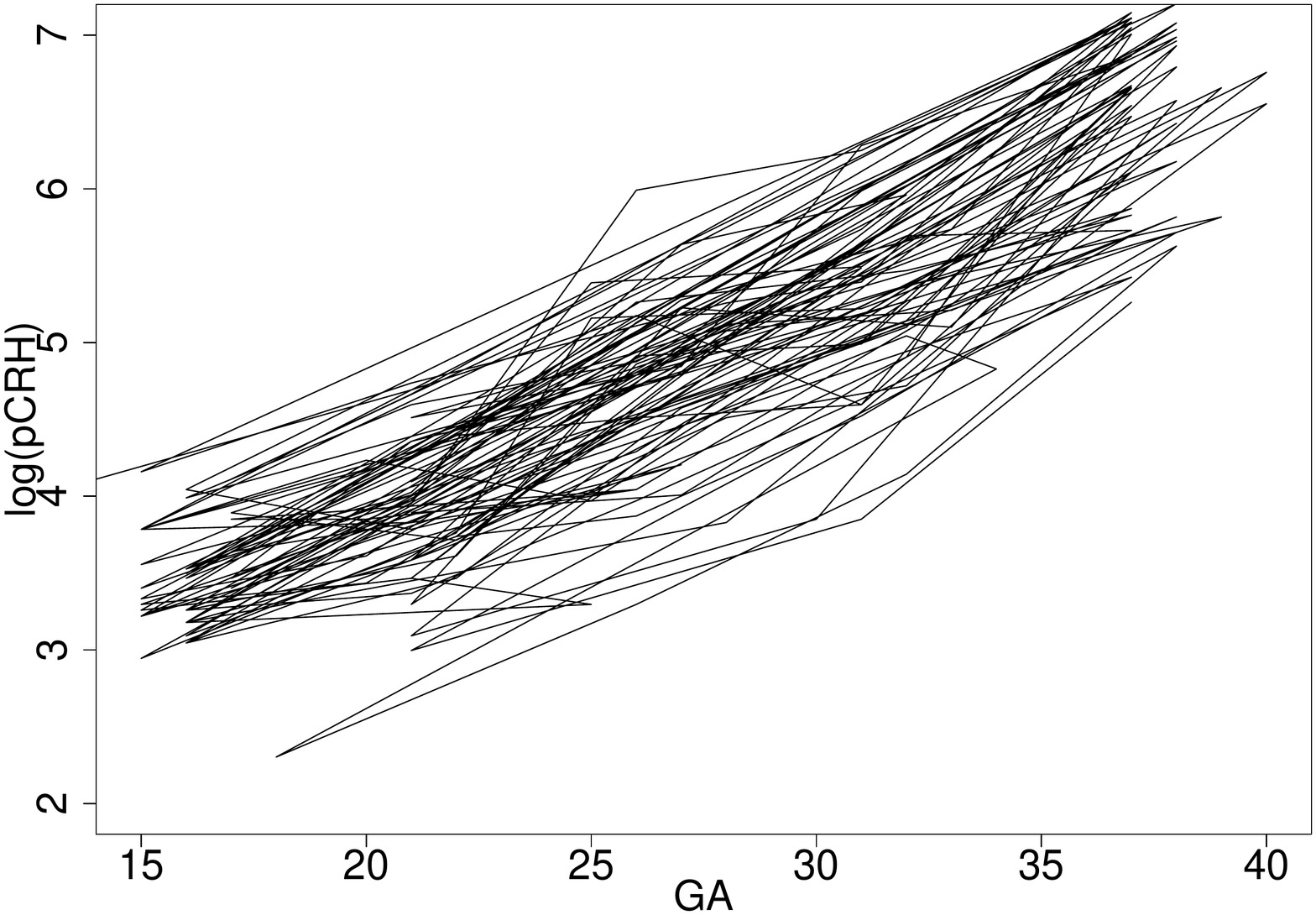} & 
                                 \includegraphics[width=.4\textwidth]{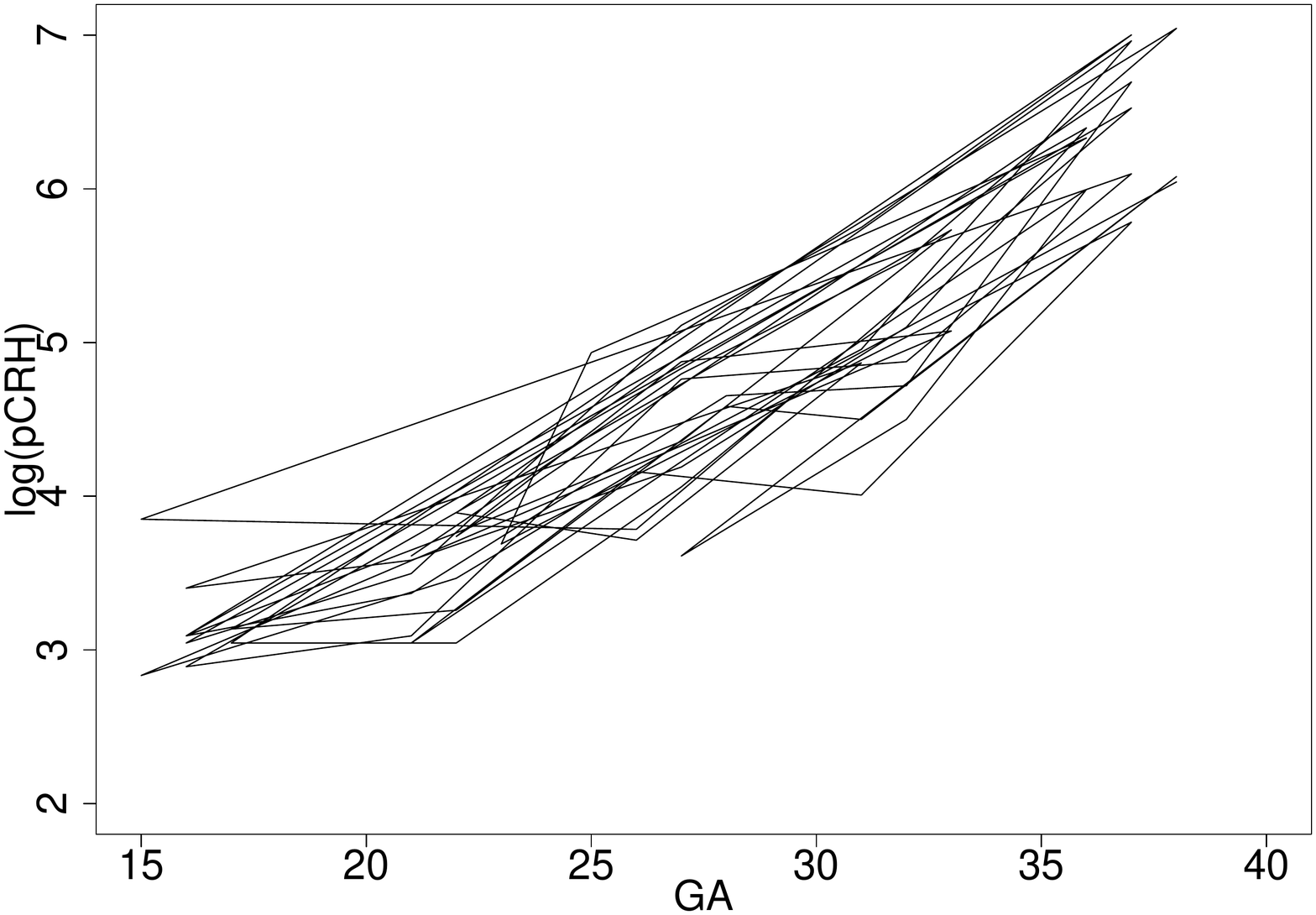} \\
                               CT-Sum=0& CT-Sum=1\\
                                 \includegraphics[width=.4\textwidth]{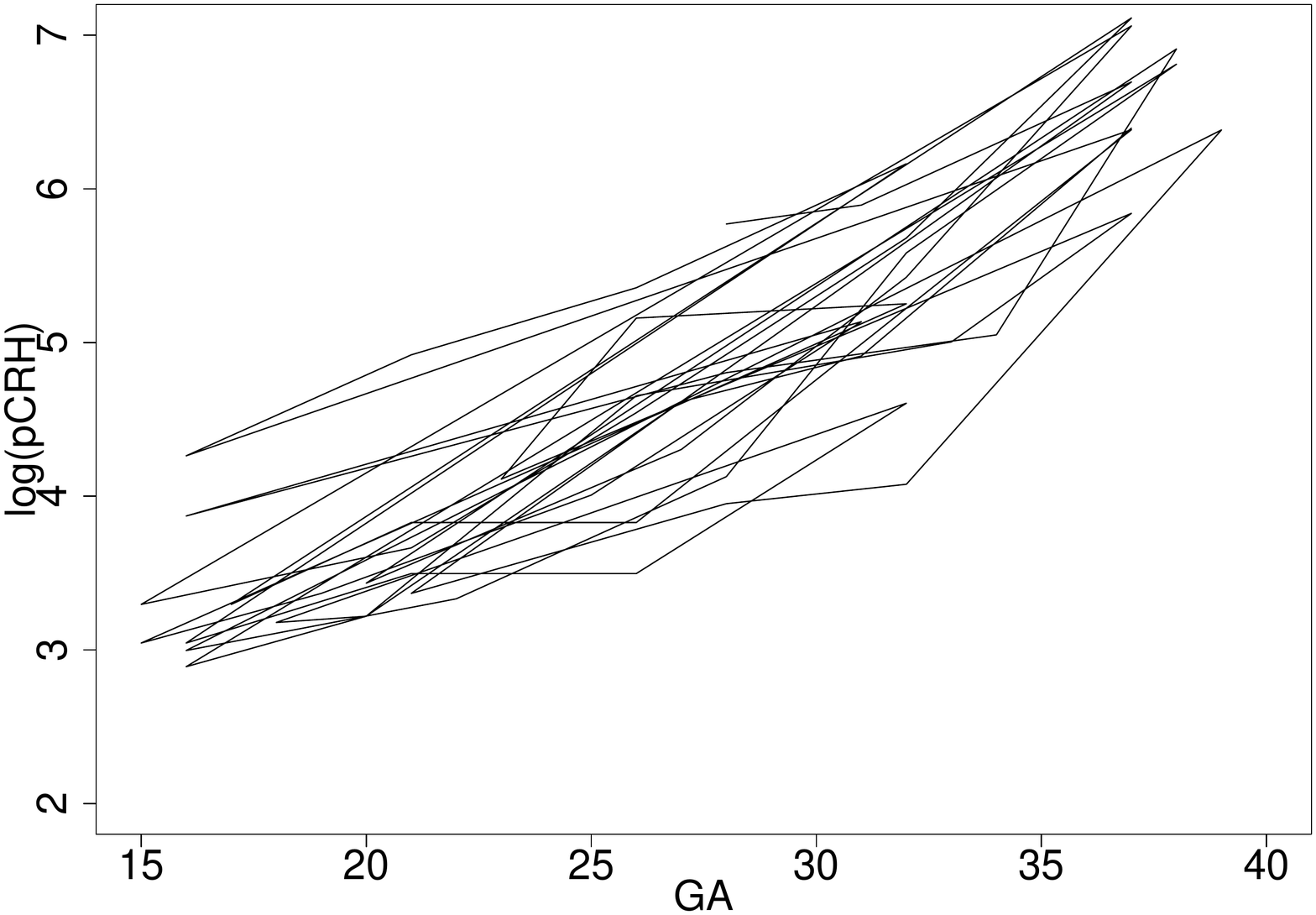} & 
                                \includegraphics[width=.4\textwidth]{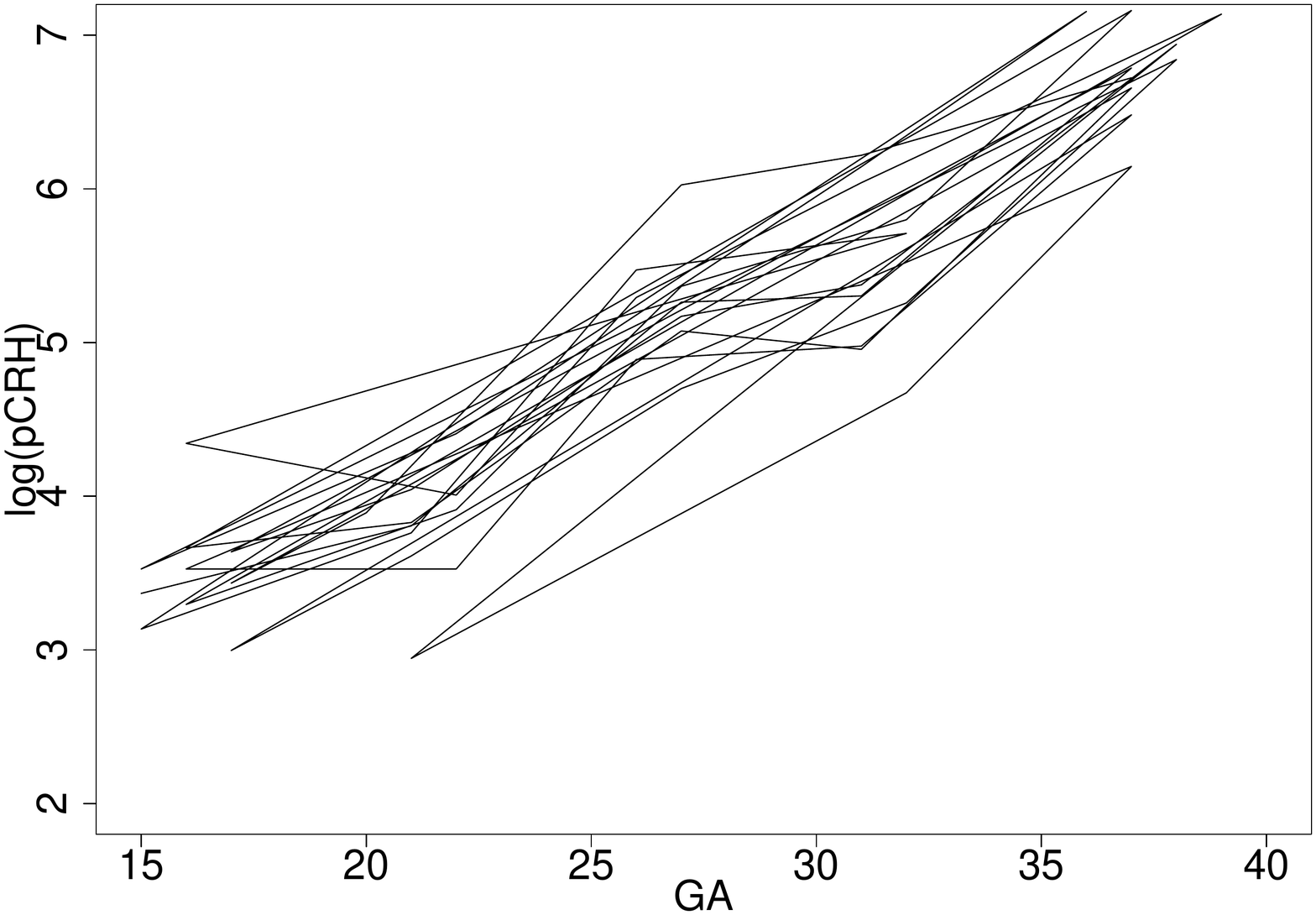} \\
                               CT-Sum=2 & CT-Sum=3\\
                                         \includegraphics[width=.4\textwidth]{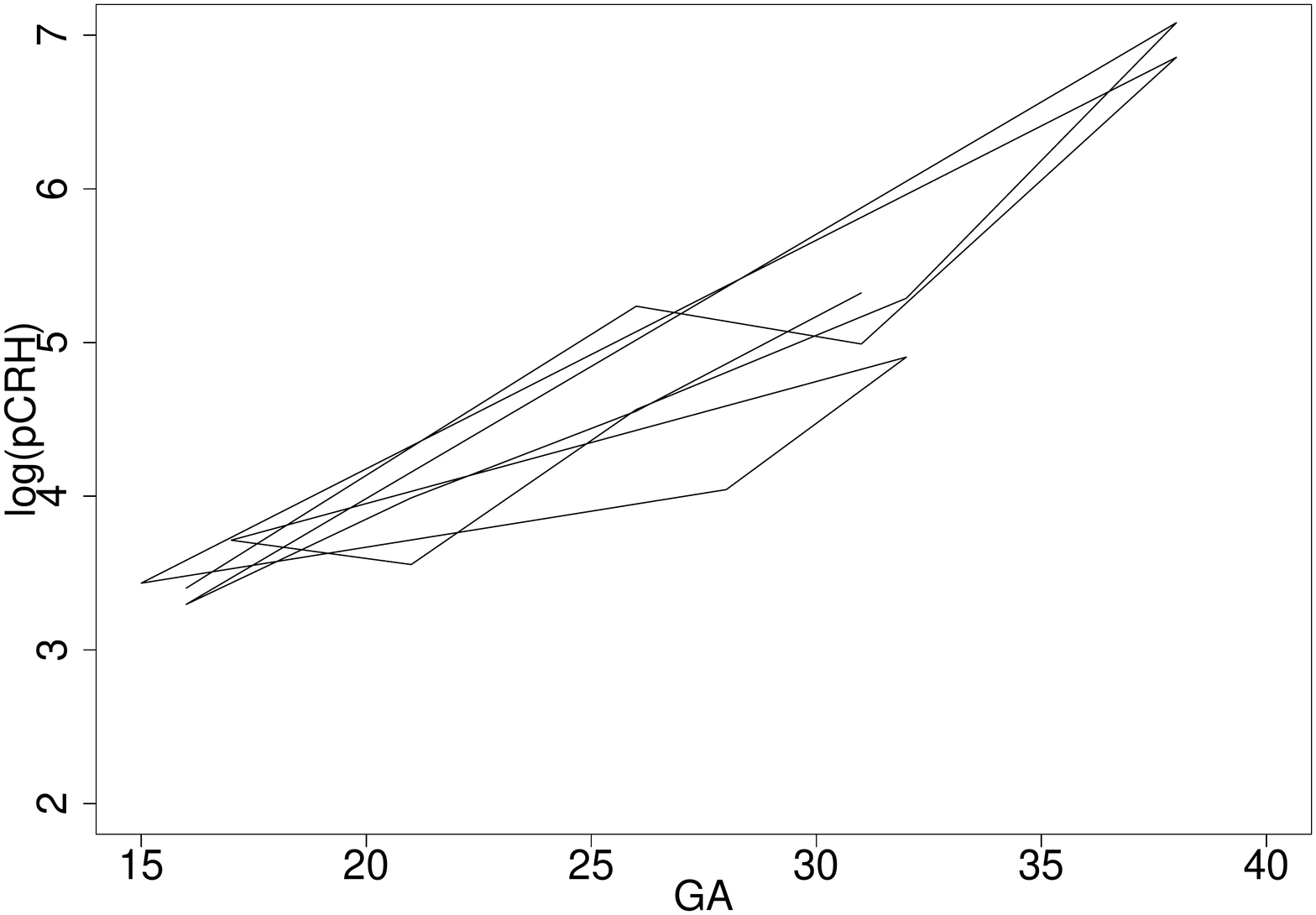} & \\
                                CT-Sum=4&
                               \end{tabular}
                               \caption{Spaghetti plots of $\log pCRH$ over GA after controlling for CT-Sum}
                               \label{trajc}
                               \end{figure}

\newpage

\section{Appendix C}
 \begin{figure}[h] \centering
                             \begin{tabular}{cc}
                               \includegraphics[width=.5\textwidth]{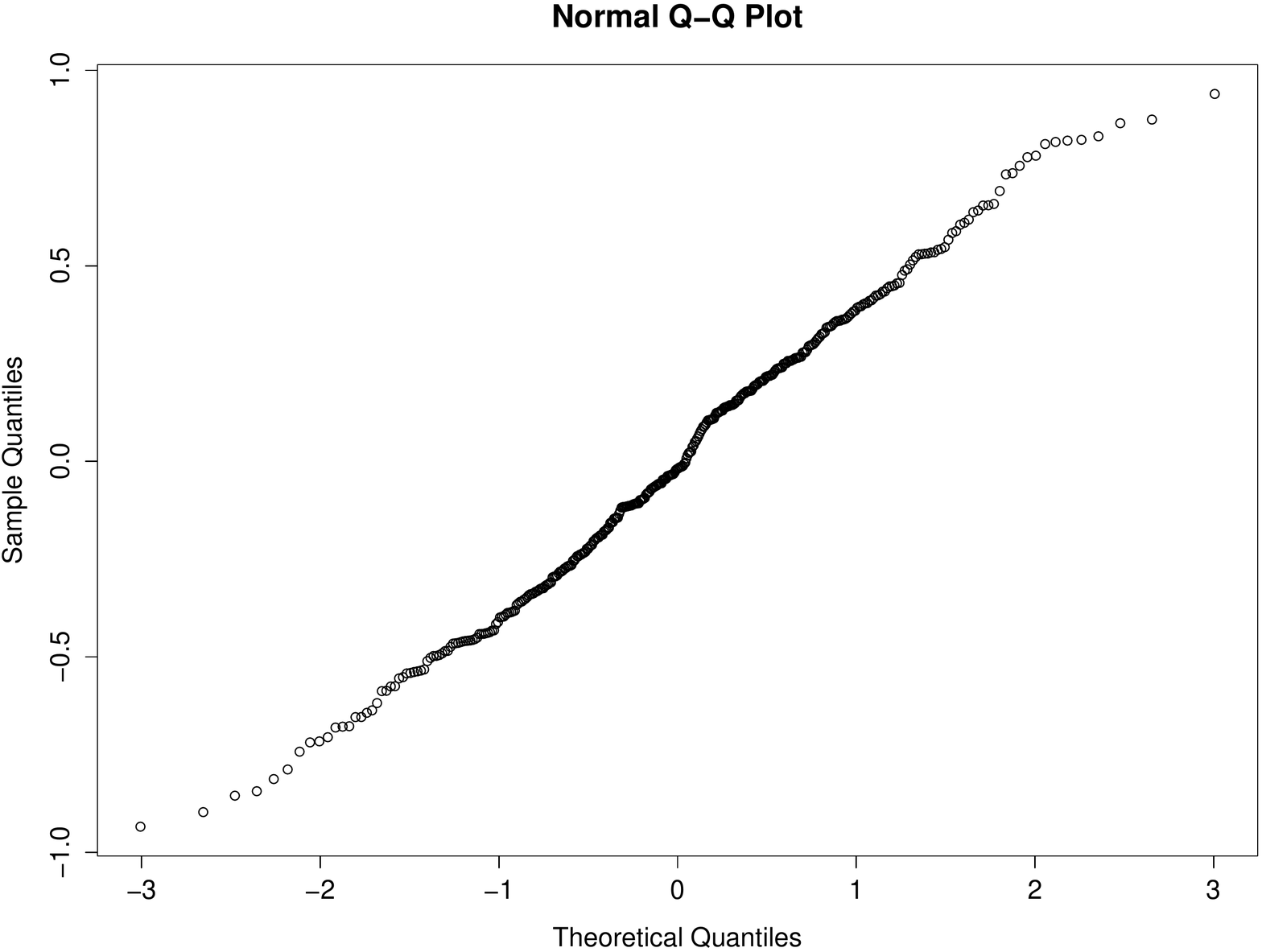} &
                               \includegraphics[width=.5\textwidth]{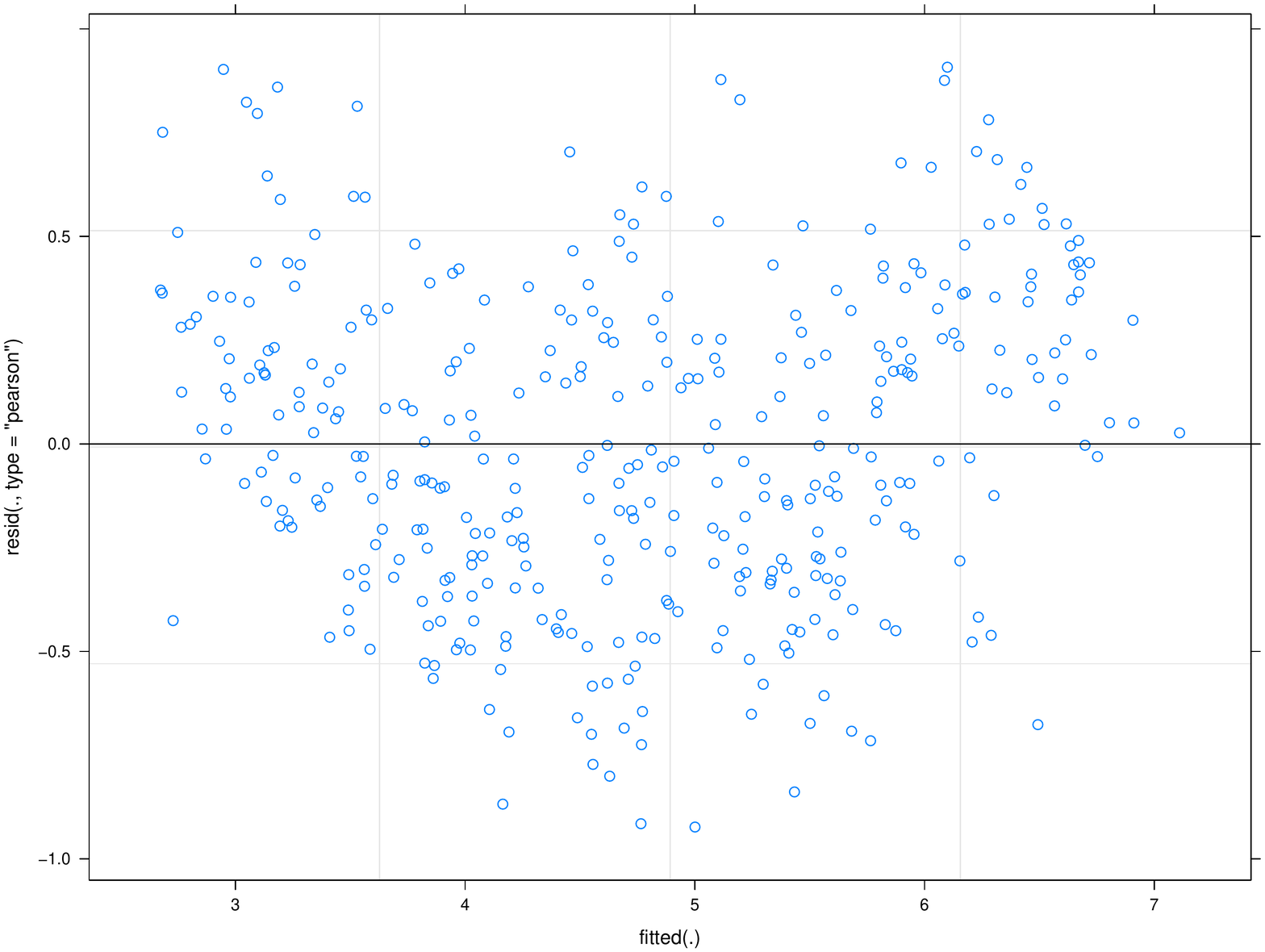} \\
                             QQ Plot &  Pearson Residual plot
                               \end{tabular}
                               \caption{QQ and Pearson plots of the model in SQ1}
                               \label{md1}
                              \end{figure}
 \newpage
 \section{Appendix D}                             
 \begin{figure}[h] \centering
                             \begin{tabular}{cc}
                               \includegraphics[width=.5\textwidth]{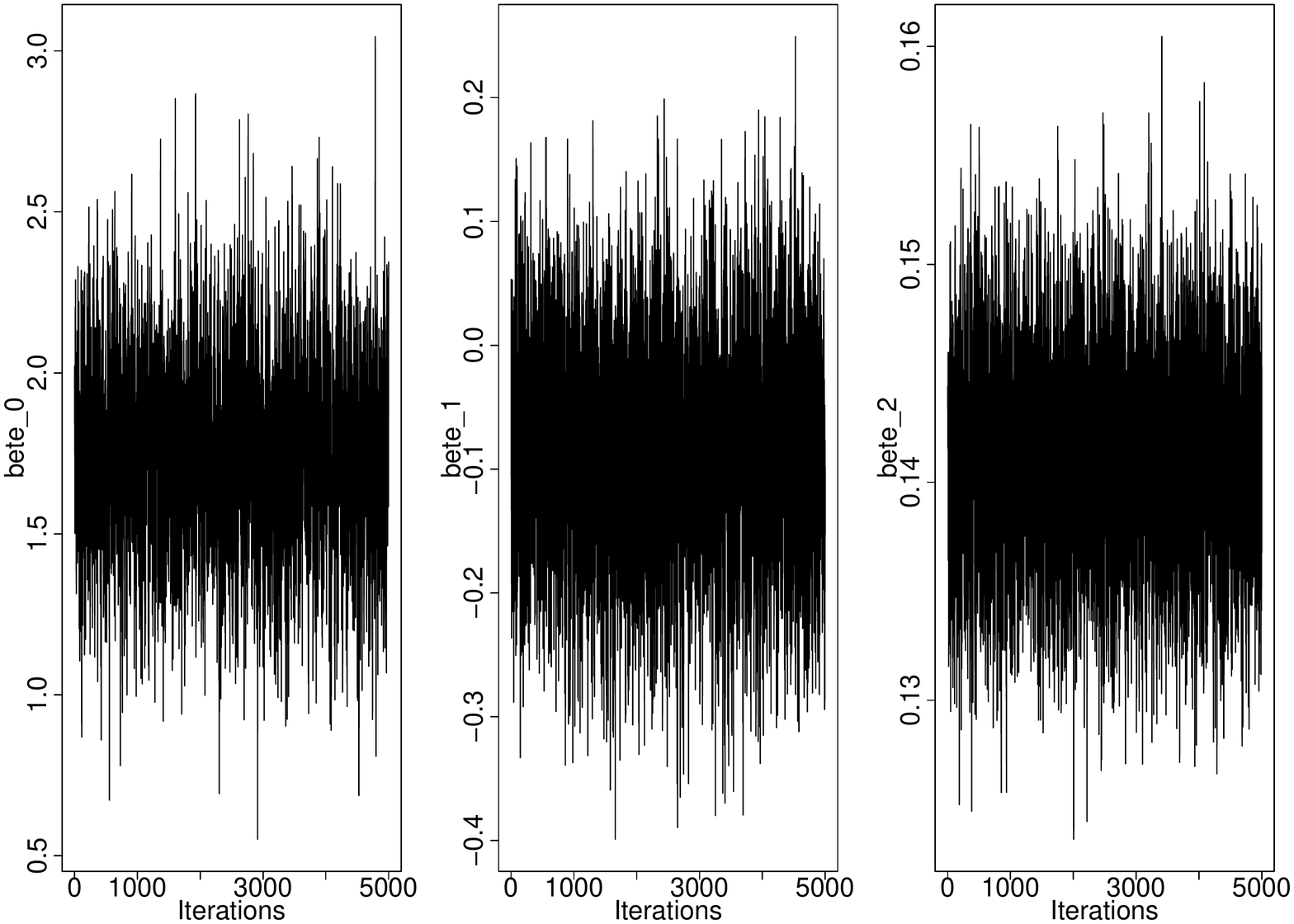} &
                               \includegraphics[width=.5\textwidth]{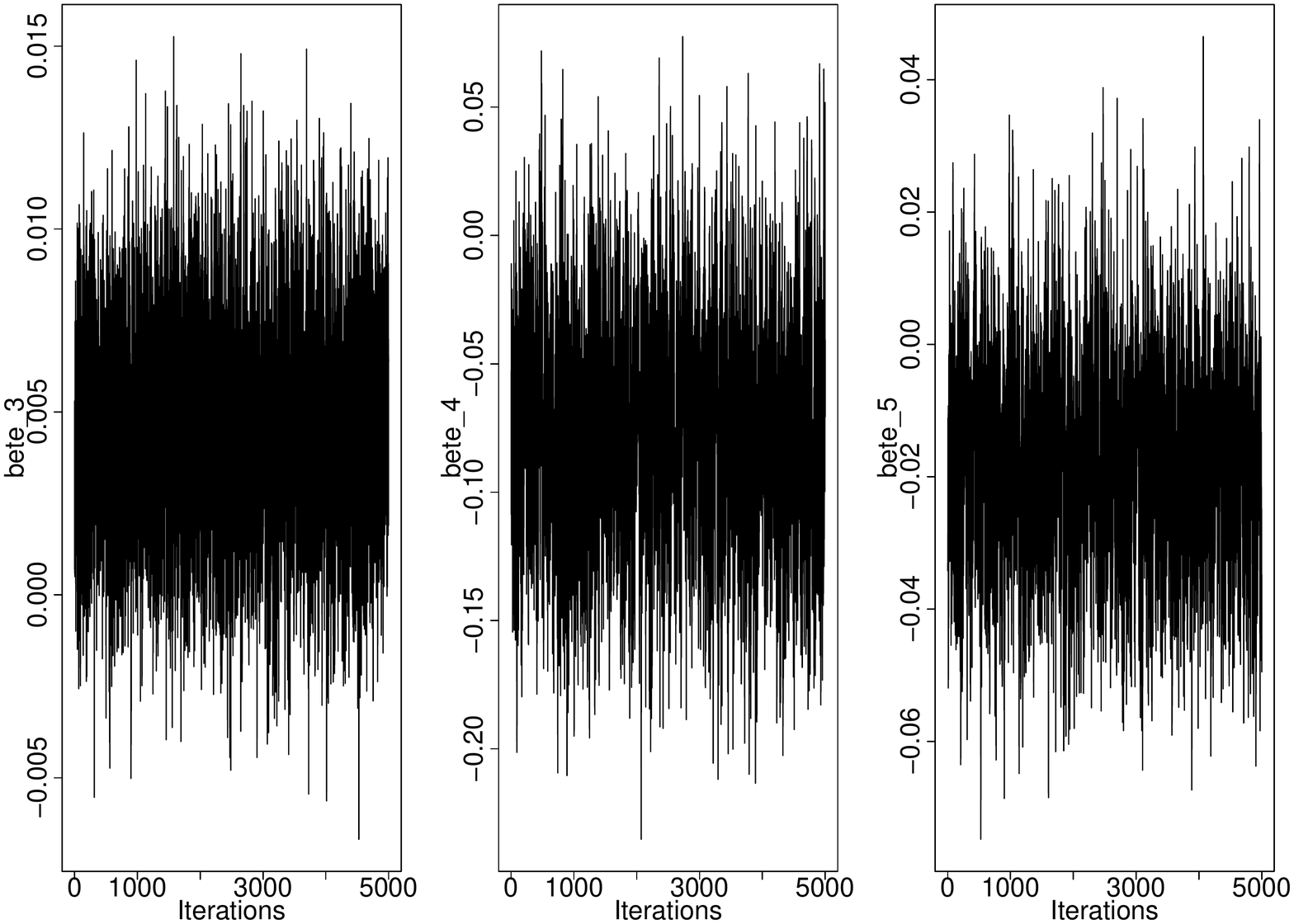}\\ 
                               \end{tabular}
                               \begin{tabular}{cc}
 \includegraphics[width=.5\textwidth]{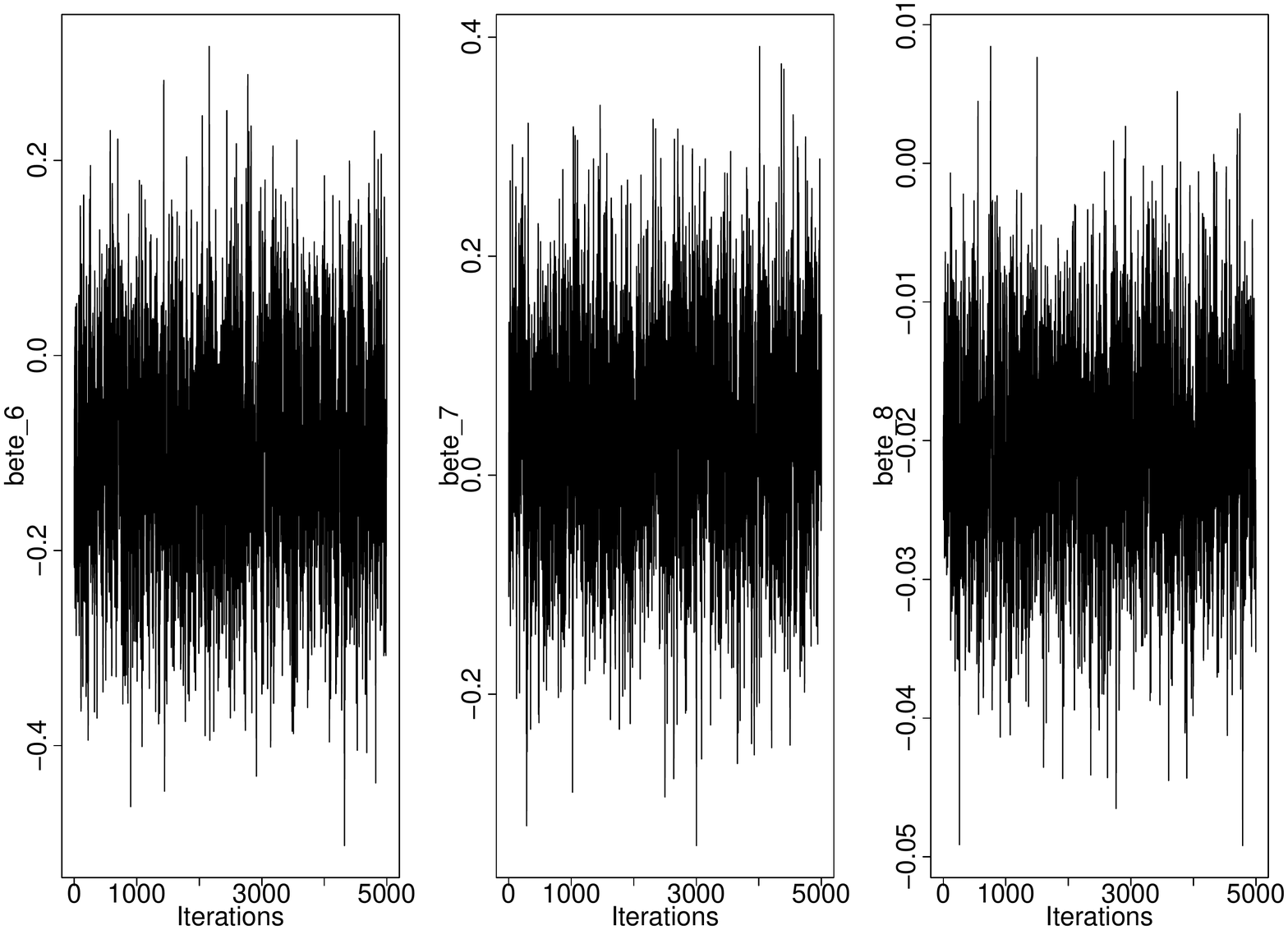} &
  \includegraphics[width=.5\textwidth]{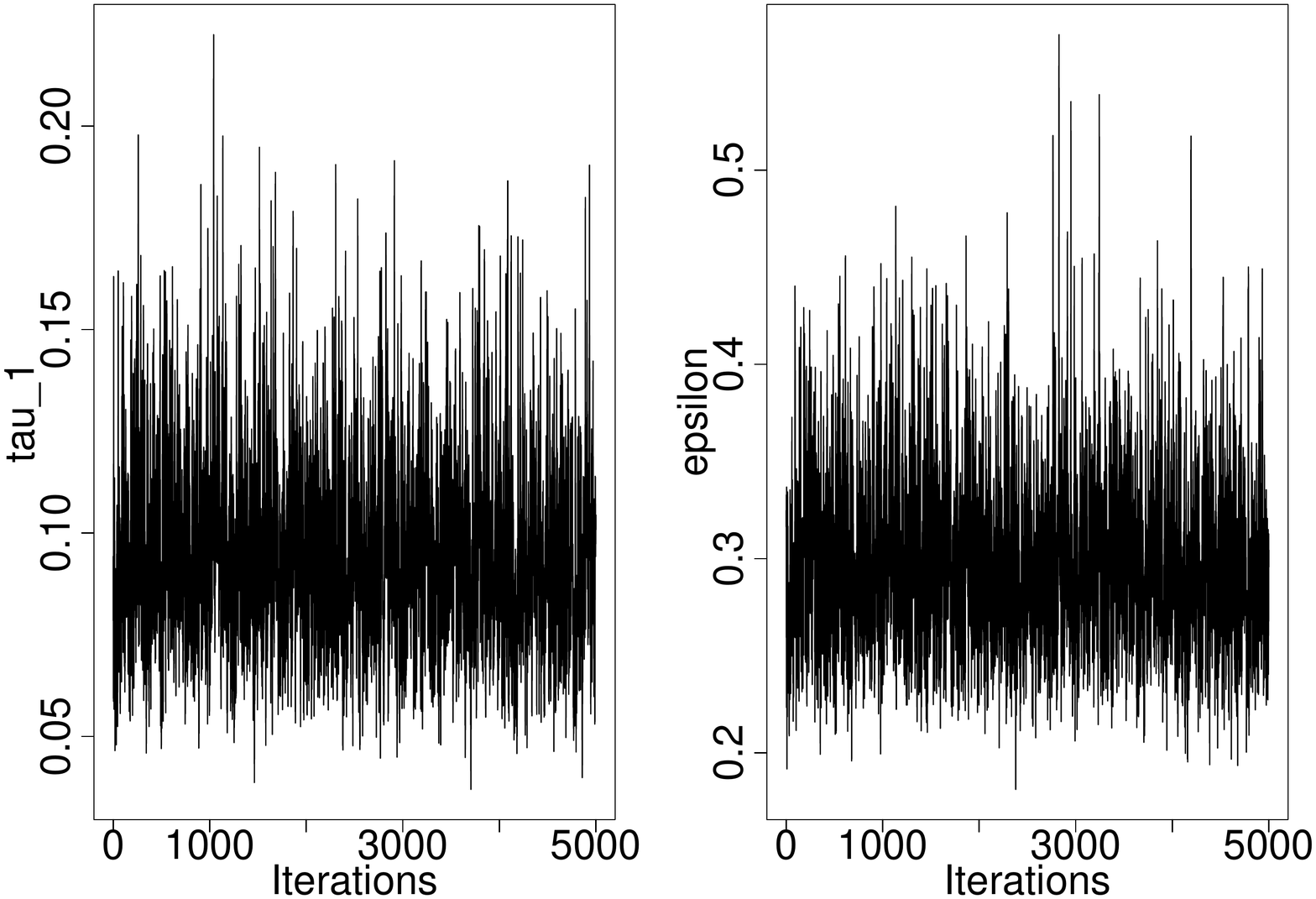}\\
 \end{tabular}
                               \caption{Trace plots of parameters of the model in SQ2}
                               \label{mdb1}
                              \end{figure}
                              
 \begin{figure}[h] \centering
                                                 \begin{tabular}{cc}
                                                   \includegraphics[width=.4\textwidth]{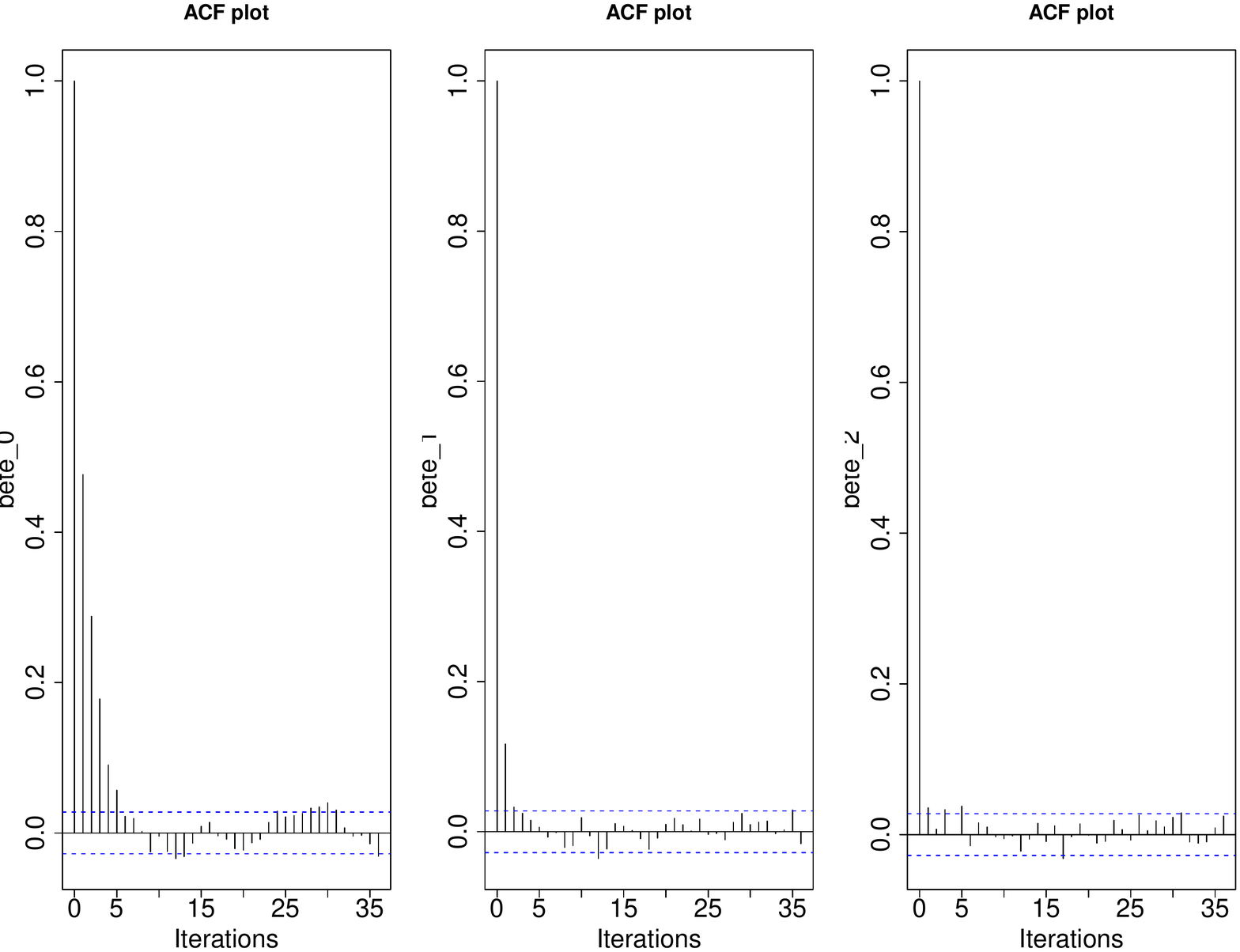} &
                                                   \includegraphics[width=.4\textwidth]{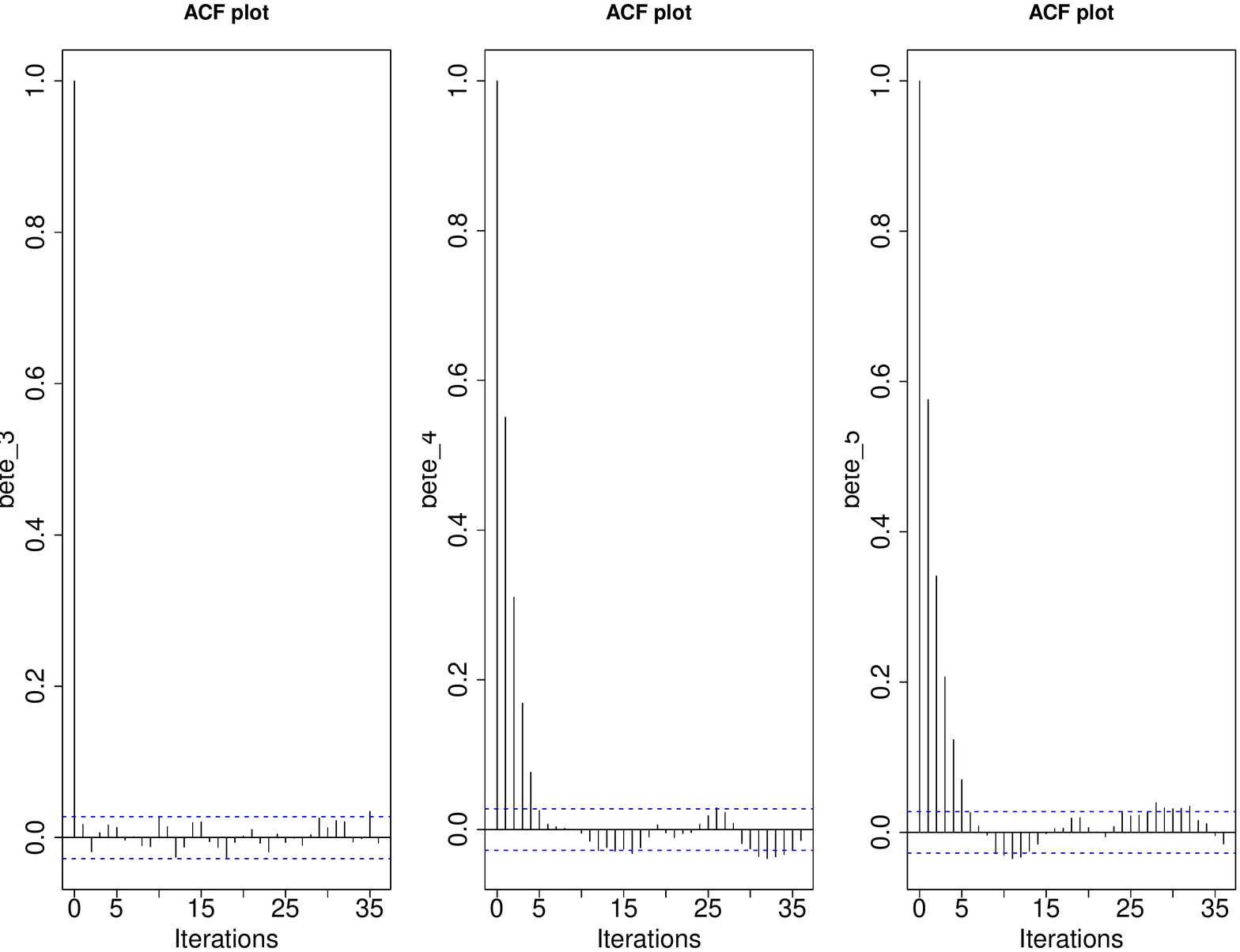}\\ 
                                                   \end{tabular}
                                                   \begin{tabular}{cc}
                     \includegraphics[width=.4\textwidth]{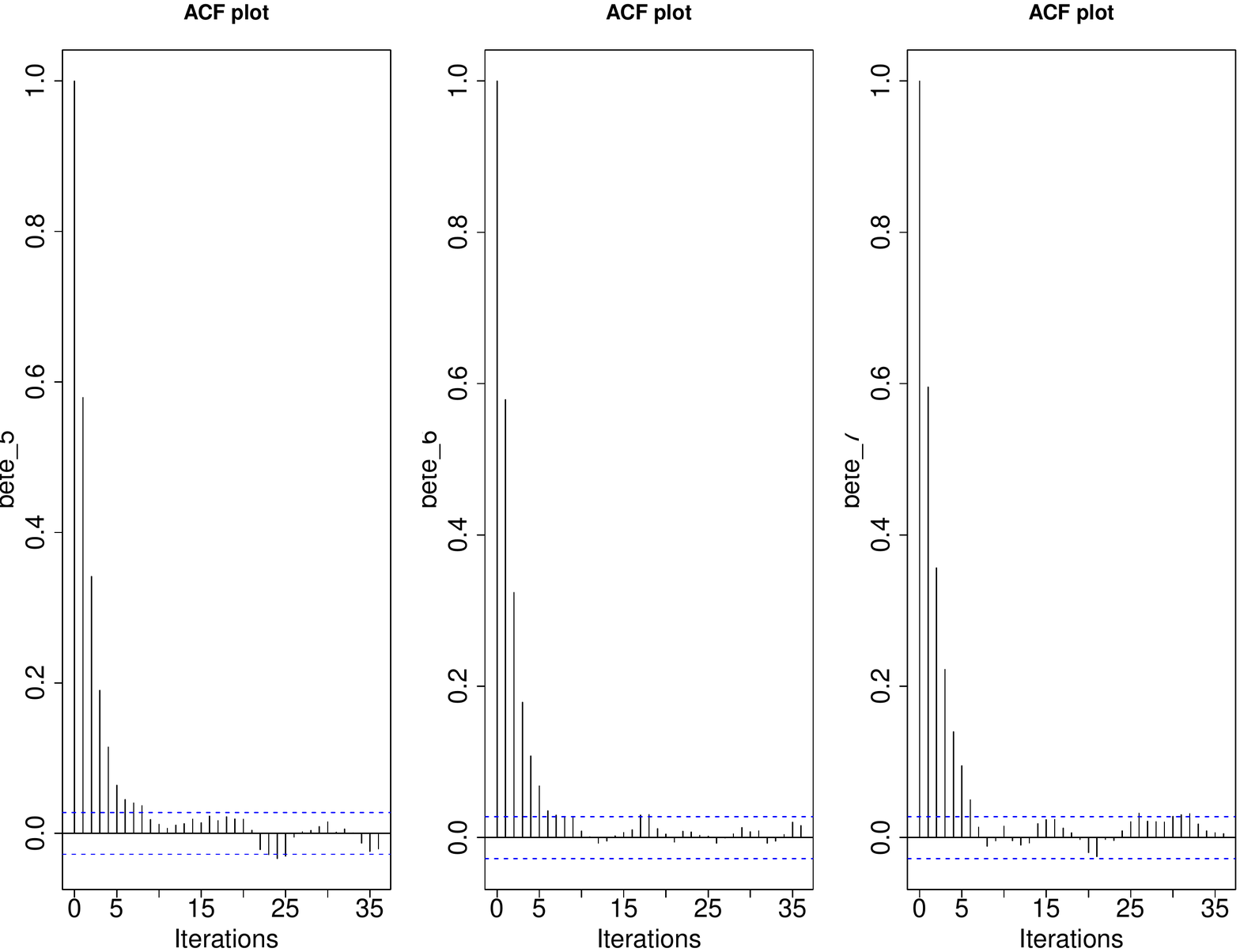} &
                       \includegraphics[width=.4\textwidth]{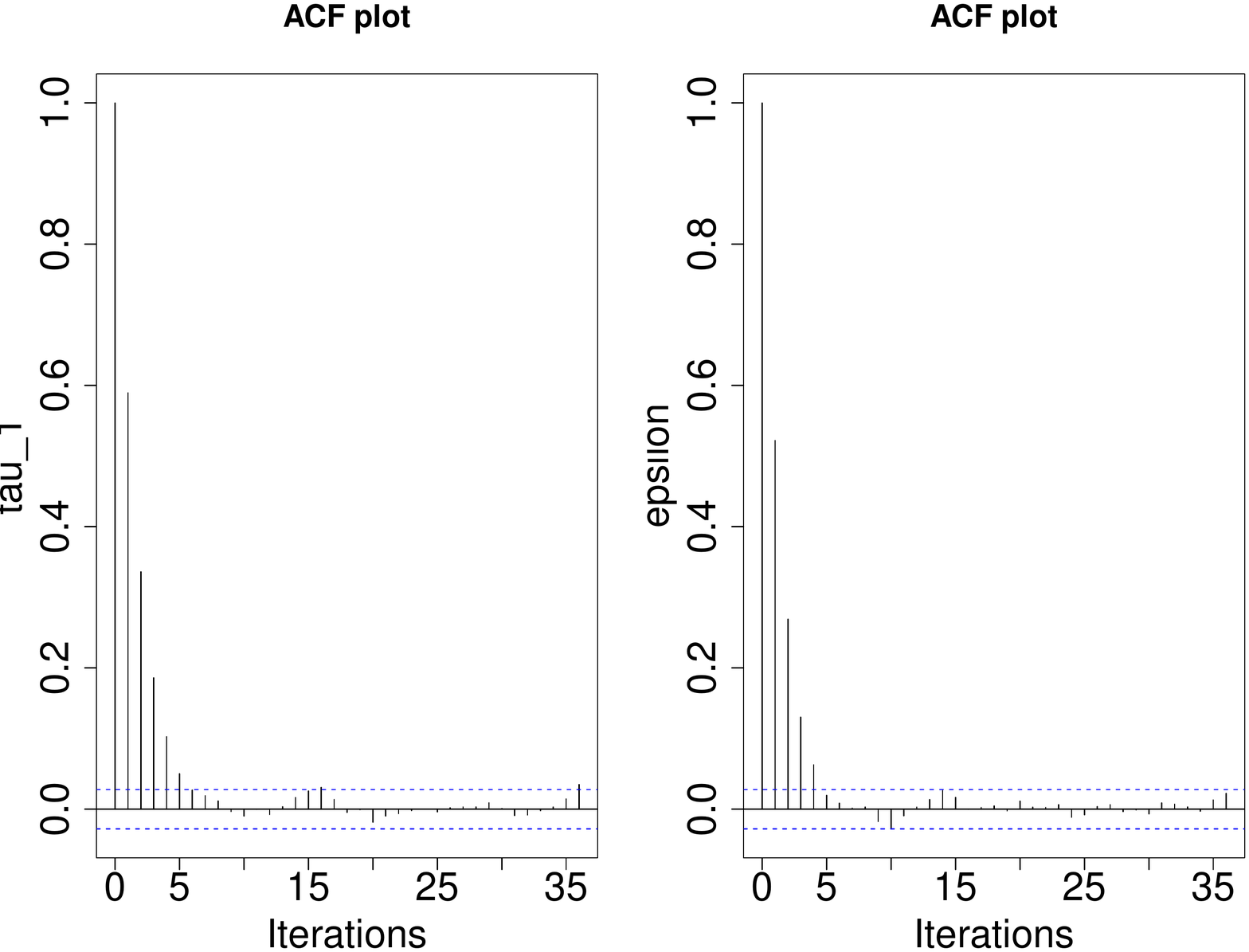} 
                     \end{tabular}
                                                   \caption{ACF plots of parameters of the model in SQ2}
                                                   \label{mdb2}
                                                  \end{figure}          
\begin{figure}[h] \centering
\includegraphics[width=.5\textwidth]{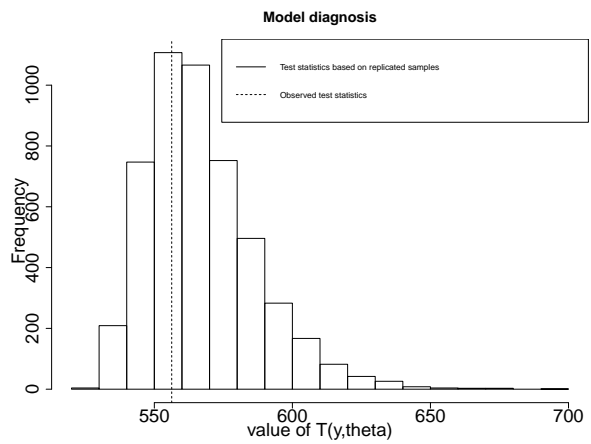} 
                              
                               \caption{Model diagnosis plot of the model in SQ2}
                               \label{md4}
                              \end{figure}                       
            \begin{figure}[h] \centering
                                        \begin{tabular}{cc}
                                          \includegraphics[width=.5\textwidth]{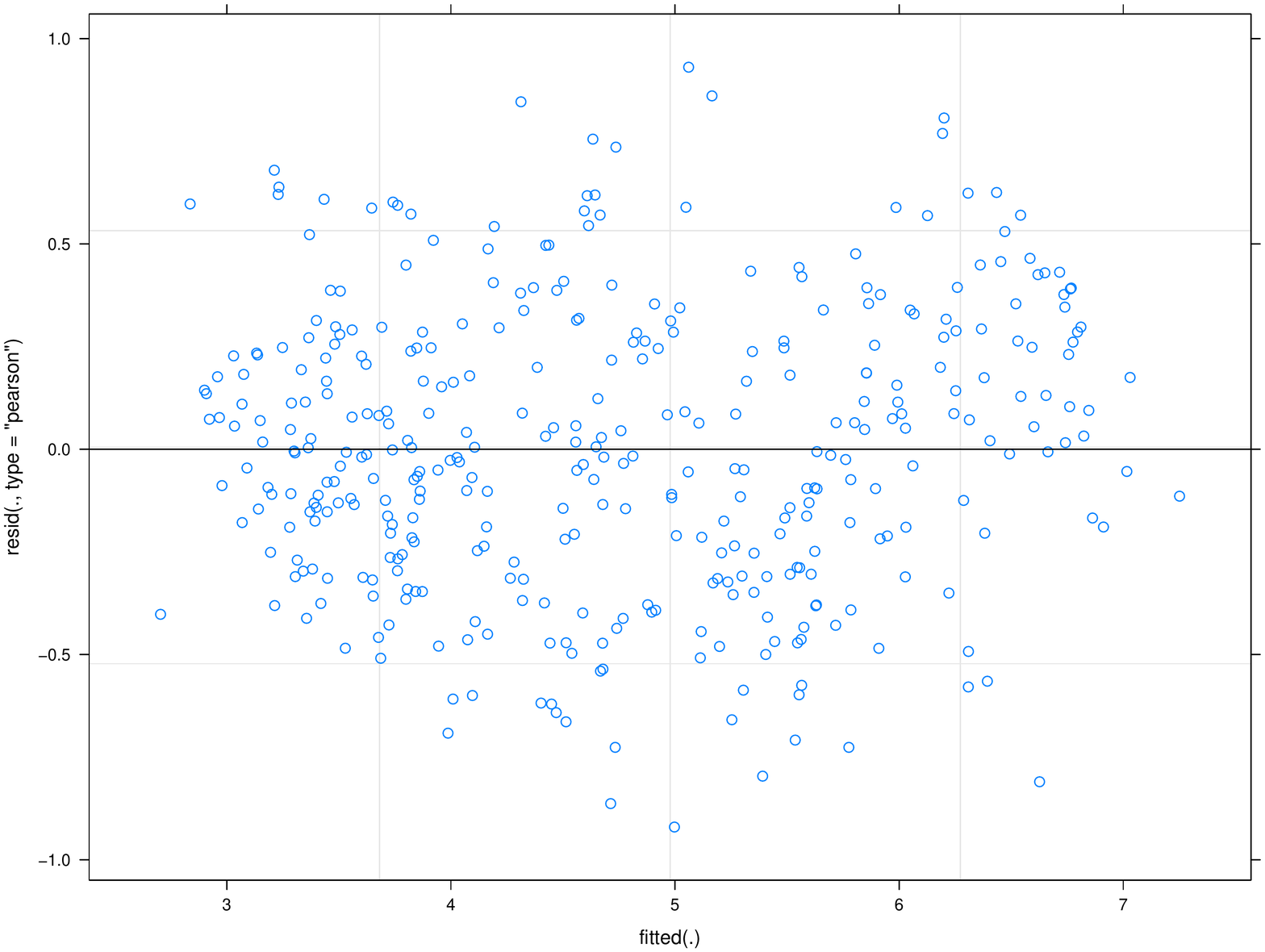} &
                                          \includegraphics[width=.5\textwidth]{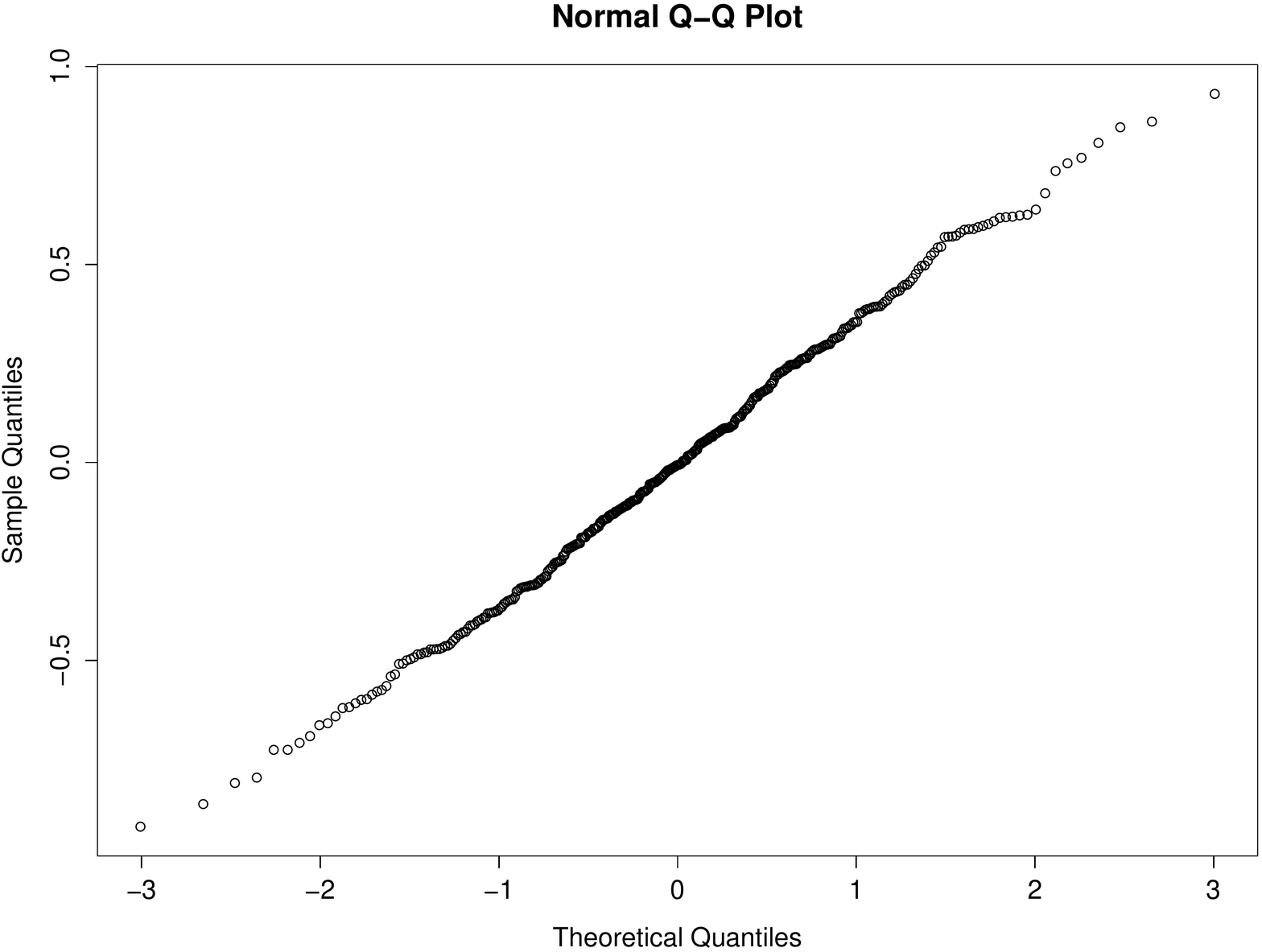} \\
                                        QQ Plot &  Pearson Residual plot
                                          \end{tabular}
                                          \caption{QQ and Pearson plots of the model in SQ3}
                                          \label{md3}
                                         \end{figure}

\end{document}